\documentstyle[preprint,prb,aps]{revtex}
\tightenlines
\begin{document}

\title{First-Principles Calculations of Hyperfine Interactions
        in La$_2$CuO$_4$}
 
\author{P. H\"usser, H. U. Suter, E. P. Stoll, and P. F. Meier}
\address{Physics Institute, University of Zurich, CH-8057 Zurich, 
         Switzerland}
\maketitle

\begin{abstract}
We present the results of first-principles cluster calculations of the electronic
structure of La$_2$CuO$_4$. Several clusters containing up to nine copper atoms
embedded in a background potential were investigated. Spin-polarized calculations
were performed both at the Hartree-Fock level and with density functional methods 
with generalized gradient corrections to the local density approximation.
The distinct results for the electronic structure obtained with these two methods
are discussed. The dependence of the electric-field gradients at the Cu and 
the O sites on the cluster size is studied and the results are compared to 
experiments. The magnetic hyperfine coupling parameters are carefully 
examined. Special attention is given to a quantitative determination of
on-site and transferred hyperfine fields. We provide a detailed analysis that
compares the hyperfine fields obtained for various cluster sizes with
results from additional calculations of spin states with different multiplicities.
From this we
conclude that hyperfine couplings are mainly transferred from nearest neighbor
Cu$^{2+}$ ions and that contributions from further distant neighbors are marginal.
The mechanisms giving rise to transfer of spin density are worked out.
Assuming conventional values for the spin-orbit coupling, the total calculated
hyperfine interaction parameters are compared to informations from experiments.
\end{abstract}


\section{Introduction}
\label{Intro}
A large amount of information about both static and dynamic properties of materials 
that exhibit high-temperature superconductivity, is available through experimental 
techniques like nuclear quadrupole and magnetic resonances (NQR and NMR). In
particular, the electric field gradients have been determined for a variety of
nuclei with great accuracy~\cite{brinkmann}. These quantities are ground-state
properties of the solid and depend sensitively on how charge is shared among the 
atomic and molecular orbitals. Measurements of the Knight shift and the various
nuclear elaxation rate tensors as a function of temperature, doping, and orientation
of the applied field with respect to the crystalline axes, provide insight into the static
spin density distribution and the low frequency spin dynamics of the electrons
in the normal as well as in the superconducting state~\cite{xyz}. Owing to the
abundances of nuclei with suitable nuclear magnetic and quadrupole moments, most
of these quantites can be studied on different sites in the unit cell. This allows
one to distinguish static and dynamic features in the CuO$_2$ planes from those in the
interlayer region.

The analysis of numerous early NQR and NMR experiments focussed on the question
whether a one-component model is sufficient to describe the low-energy excitations
of the quantum fluid in the high-temperature superconductors or whether a two-component
description is needed. Knight shift and
relaxation rate data for the Cu sites in YBa$_2$Cu$_3$O$_7$ that were analyzed
on the basis of the hyperfine coupling of a Cu$^{2+}$ ion in an axial symmetric
surroundig, seemed to be controversial. Mila and Rice~\cite{milarice} proposed
then that besides the usual on-site hyperfine interactions at a Cu nucleus an additional 
hyperfine field should be considered which is transferred from neighboring Cu ions.
They gave a consistent explanation of the unusual combination of anisotropies of 
the Cu Knight shifts and relaxation  rates in YBa$_2$Cu$_3$O$_7$ within a one-component 
theory. An estimation of the strength of the various hyperfine parameters with the
use of a quantum chemical model led to good agreement with the values deduced
from experiment. Monien, Pines, and Slichter~\cite{mps} analyzed a more complete
set of data from which they inferred the relevant hyperfine couplings and the associated
fluctuation times. The ensuing constraints on theoretical one-component models
for the planar magnetic excitations comprise a large transferred hyperfine field
between adjacent Cu$^{2+}$ spins, which, moreover, must be very nearly aligned
antiferromagnetically, and, in addition, an almost perfect cancellation between 
the parallel component of the dipolar hyperfine tensor and the total transferred
field $(A_{\parallel} + 4 B \approx 0)$. Provided that the spin susceptibility is
isotropic, it is only with the latter condition that the vanishing of the parallel
component of the spin contributions to the Knight shift can be explained.

A remaining intricate problem whose explanation seemed to require a two-band model,
was the relaxation behavior of the planar oxygen nuclei which was strikingly different
from that of the coppers. Shastry~\cite{shastry}, however, pointed out that spin density
is also transferred to the planar oxygens. A transferred hyperfine field couples a
nucleus to spin fluctuations on neighboring Cu sites. This implies that the spin-lattice
relaxation time $T_1$ depends on form factors given by the local geometry of lattice sites
of the nucleus and its neighbors. Since the form factor entering $^{17}T_1$ vanishes at
the commensurate antiferromagnetic wave vector $(\pi/a,\pi/a)$ where that for  $^{63}T_1$
reaches a maximum, the variant temperature dependence of  $^{17}T_1$ and  $^{63}T_1$
could be understood from the presence of strong antiferromagnetic correlations between
the copper spins in the planes. Thus a coherent picture of the low-energy spin dynamics
in the CuO$_2$ planes seemed to be established~\cite{ww} and numerous NMR and NQR experiments
were analyzed on the basis of these hyperfine spin Hamiltonians and a one-band model.
Pines and coworkers developed~\cite{mmp} a phenomenological expression for the dynamic
spin susceptibility and provided~\cite{moni,bp} a quantitative description of a variety of NMR data.

New doubts about the adequacy of a one-component description of the spin fluid
arose when the results of neutron scattering data became available. NMR and neutrons both probe
the spin susceptibility. In La$_{1.86}$Sr$_{0.14}$CuO$_4$, 
a considerable degree of incommensuration of the peaks in the spin-fluctuation spectrum 
was observed~\cite{aeppli}. As a consequence, the NMR relaxation time $^{17}T_1$ is
expected to exhibit an anomalous temperature dependence which is not seen experimentally.
This and other contradictions between the results of NMR and neutron-scattering
experiments led Zha, Barzykin, and Pines~\cite{zha} to a critical
re-examination of the hyperfine spin Hamiltonian. They showed that it is possible to reconcile NMR and
neutron-scattering experiments on both La$_{2-x}$Sr$_x$CuO$_4$ and
YBa$_2$Cu$_3$O$_{6+x}$ within a one-component theory by introducing a transferred
hyperfine coupling between the O nuclei and their next-nearest-neighbor Cu$^{2+}$
spins. In addition, the analysis exhibited that the hyperfine coupling transferred
to the Cu nuclei changes as one goes from the  La$_{2-x}$Sr$_x$CuO$_4$ to the
YBa$_2$Cu$_3$O$_{6+x}$ system, and is moreover comparatively sensitive to hole
doping in the former system.

Despite the rich information of NMR and NQR results about the nature of the spin
fluid and the low-energy quasiparticles, there exist only few theoretical
first-principles approaches which address the determination of electric-field
gradients and magnetic hyperfine interactions. For the YBa$_2$Cu$_3$O$_{6+x}$ 
system, electric-field gradients (EFGs) at the various nuclear sites have been obtained by Das and 
coworkers~\cite{das,das1,das2} with {\it ab-initio} cluster calculations using the
unrestricted Hartree-Fock (UHF) method, and by Schwarz and 
coworkers~\cite{schwarz1,schwarz2} who employed the full-potential 
linear augmented-plane-wave method within the local density approximation (LDA).
Results of Hartree-Fock (HF) calculations have also been published by
Winter~\cite{winter}. Apart from one exception, the EFGs calculated with these
three different methods more or less agree and reproduce the experimental
data quite satisfactorily. The exceptional case is the EFG at the planar Cu
sites. Using a large cluster that contained 74 atoms, we recently 
performed~\cite{huesser} calculations with the density functional (DF) method and
obtained values for the Cu EFGs which are in better agreement with the experiments.
These calculations were non-spin polarized and no information on magnetic 
hyperfine interactions is available. The energy difference between the singlet and
triplet states of two coppers in adjacent CuO$_2$ planes was studied~\cite{suter1}
in a small cluster containing 2 Cu, 4 Y, and 8 O atoms. 

For the La$_{2}$CuO$_4$ system, UHF calculations have been reported by 
Sulaiman et al.~\cite{sulaiman} who obtained EFG values at various nuclei as 
well as the
contact and dipolar hyperfine fields at Cu. Martin and Hay~\cite{martinhay} presented
HF results of the electronic structure of CuO$_6$ clusters in neutral, p-doped,
and n-doped states and studied the influence of correlation effects
using the technique of configuration interactions. Martin~\cite{martin}
calculated HF values for EFGs in CuO$_6$ and Cu$_2$O$_{11}$ clusters and
investigated the change in the NQR spectra upon doping.
A comparison~\cite{suter2} of HF and LDA calculations for a CuO$_6$ cluster
showed large differences in the EFG values and the hyperfine coupling parameters.

In this paper, the results of extensive cluster studies for the pure La$_2$CuO$_4$
system are presented. Spin-polarized calculations at the HF level and with 
the DF method with gradient corrections to the correlation functionals have
been performed for clusters comprising n = 1, 2, 4, 5, and 9 Cu atoms in a plane.
The resulting electronic structures, the charge and spin distributions, EFGs 
and magnetic hyperfine interactions have been analyzed.
In Sec.~\ref{clusters}, details on the clusters, their embedding in a 
lattice of point charges, the basis sets and the theoretical methods are given.
Results for the CuO$_6^{10-}$ ion are presented in Sec.~\ref{Ion} and compared to the
predictions for the hyperfine couplings of a Cu$^{2+}$ ion in a single-electron
picture. The distinct results of HF and DF methods are discussed in detail.
Section~\ref{efg} is devoted to the investigation of the EFG values for Cu
and the planar and apical oxygens and their dependence on the cluster size. 
The results for the magnetic hyperfine interactions are given in
Sec.~\ref{hff}. Particular emphasis is put on investigations of contributions to the
hyperfine fields transferred from nearest and further distant copper ions.
The origins of the various hyperfine couplings are worked out and explained in terms of
localized atomic orbitals and molecular orbitals. The general mechanisms giving rise to
spin density transfer are discussed.
Sec.~\ref{con} contains a summary and conclusions.

\section{Clusters and Basis Sets}
\label{clusters}
The general idea of the cluster approach to electronic structure calculations 
is that the parameters that characterize a small cluster should be transferable to the
solid and largely determine its properties. The essential contributions to EFGs and
to magnetic hyperfine fields are given by rather localized interactions and
therefore it is expected that these local properties can be determined and understood
with clusters calculations. Approximations must be made concerning the treatment of the
background that is employed to embed the cluster. Using as large a cluster as is
possible is of course advantageous. It is necessary, however, that the results
obtained should be checked with respect to their dependence on the cluster size.

The clusters used in this work comprise copper atoms of the CuO$_2$ plane together
with the surrounding planar and apical oxygens. They were embedded in a lattice
of point charges with charges $+2, -2,$ and $+3$ according to the ionic charges
of Cu$^{2+}$, O$^{2-}$, and La$^{3+}$, respectively. However, it has been pointed 
out in Ref.~\cite{martinhay} that bare positive point charges close to the border
of the cluster present a too strong attraction for the electrons. It is therefore
essential to replace them by appropriate pseudopotentials which guarantee the
orthogonality of the diffuse electron wave-functions with the ion cores.

The smallest cluster investigated is shown in Fig.~\ref{clustercu1}.
All-electron basis sets were put on one Cu and six O atoms and pseudopotentials
were used to represent the four Cu$^{2+}$ and ten La$^{3+}$ sites closest to
the cluster border. In addition, these 21 atoms were surrounded by more than 2000
point charges. Some charges at sites far from the cluster center were adjusted
such that the total charge of the system was zero. Next, the positions of some
point charges were changed in such a way that the correct Madelung
potential in the central region of the cluster was reproduced.

All atomic positions were located according to the tetragonal phase with
lattice constants~\cite{lattice} a = 3.770 {\AA} and c = 13.18 {\AA} and with 
a Cu-O(a) distance of 2.40 {\AA}.

For the cluster atoms the standard 6-311 G basis sets were employed. For Cu, this
corresponds to the basis set developed by Wachters~\cite{Wachters}, while
for O it was given in Ref.~\cite{pople}.
In Table~\ref{clust} an overview is presented of the 
clusters used in this work with a listing of the numbers of Cu atoms, of cluster atoms,
of electrons, of basis functions, and of primitive Gaussian functions.

The majority of the HF and DF calculations were spin-polarized. 
They were performed with the Gaussian94 and Gaussian98 programs \cite{g98}.
The Vosko-Wilk-Nussair functional (VWN)~\cite{vwn1980} was used as local density
approximation (LDA). For the gradient corrections to the exchange and correlation
functionals several forms for the generalized gradient approximations (GGA) are 
available in the literature. As in our investigations~\cite{huesser} on YBa$_2$Cu$_3$O$_7$, we 
mainly used the formula proposed by Becke~\cite{becke1,becke2}
in combination with the functional of Lee, Yang, and Parr~\cite{LYP} (BLYP).
For the smallest cluster, other functionals have been used as well, such as
the X$_{\alpha}$-LYP (XALYP) and the form proposed by Perdew and Wang~\cite{PW91}~(BPW91).
The individual integrals over the atomic orbitals for the electric 
field gradient were taken from calculations using the GAMESS--US program~\cite{gamessus}.
For the final analysis a special program was designed~\cite{stoll97} which employed 
these integrals and the electronic wave functions from the Gaussian calculations .

Throughout this work we use atomic units for the electric-field gradients and
magnetic hyperfine interaction parameters. To distinguish 
these from other notations, we will use lower case letters and write, e.g., the
hyperfine spin Hamiltonian for a nuclear spin $I$ and a single electron spin
$S$ as
\begin{equation}
	H = \sum_{\alpha\beta} \; I_{\alpha} \; A_{\alpha\beta} \; S_{\beta}
		= \hbar \gamma_I \hbar \gamma_e \, 
		\sum_{\alpha\beta} \; I_{\alpha} \; a_{\alpha\beta} \; S_{\beta} \quad .
\end{equation}
The tensor $a_{\alpha\beta}$ has the dimension of a density. In the NMR literature, 
different and sometimes misleading units are in use. We note that 
$2 \mu_B /a_B^3 = 125.2$ kG, which connects to the units used in Ref.~\cite{mps}.
\section{Embedded C{u}O$_6$-Ion}
\label{Ion}
To investigate the (CuO$_6$)$^{-10}$ ion embedded in an 
environment of the La$_2$CuO$_4$ system, the cluster CuO$_6$/Cu$_4$La$_{10}$
shown in Fig.~\ref{clustercu1} was used. The results for the EFG 
component $V_{zz}$, the core
polarization, the dipole tensor, and the atomic spin density $\rho$ at the
Cu site are collected in Table~\ref{cuo6}.

Our results for $V_{zz}$ and ${\rm a_{dip}}$
obtained at the HF level agree with previous predictions.\cite{sulaiman,martin} 
There exist, however, distinct differences between the HF results
and those calculated within the DF/GGA scheme.
In HF, the EFGs are about 40 \% larger than those obtained from DF calculations.
A similar discrepancy between EFG 
values calculated with HF and DF methods has also been observed~\cite{huesser} in 
calculations of the EFG at the planar Cu site in the 
YBa$_{2}$Cu$_{3}$O$_{7}$ system.

The results of different calculations within the DF approach, performed by using
other exchange and correlation functionals (SVWN, XALYP, BPW91), however,
agree among each other reasonably well.  
Increasing the quality of the basis set  by including
polarization (p) and diffuse (d) functions also does not alter the results 
substantially.

The experimental value~\cite{imaisl} of the quadrupole frequency at the $^{63}$Cu 
is 33.0 MHz. This corresponds to $a_{zz} = 1.331$ assuming
a quadrupole moment $Q(^{63}$Cu) = $-.211$ b. A more detailed comparison between
experimental and theoretical values for the EFG is given in Sec.~\ref{efg}.
The results for the core polarization and the dipole tensor will be discussed
and compared to experiments in Sec.~\ref{hff}.

It is instructive to compare these {\it ab-initio} results for the embedded
(CuO$_6$)$^{-10}$ ion with the predictions for a single-electron 
(or rather single-hole) picture of the Cu$^{2+}$ ion which have been
given by Bleaney et al.\cite{Bleaney}. If the d-shell is full except for a
singly occupied 3d$_{x^2-y^2}$ orbital, the dipolar hyperfine field is given by

\begin{equation}
	  {\rm a^{\parallel}_{dip}} =   - \frac{4}{7} < r^{-3} > 
\end{equation}
and the core polarization is determined by a parameter $\kappa$:

\begin{equation}
        {\rm a_{cp}}     =    - \kappa   < r^{-3} >                \, .
\end{equation}
The spin-orbit coupling will be discussed later.

Inserting the DF-value ${\rm a^{\parallel}_{dip} = -3.526}$ from Table~\ref{cuo6} 
into Eq. (1)
gives $< r^{-3} >$ = 6.171. Using this number and  ${\rm a_{cp} = -1.784}$ in Eq. (2) 
leads to $\kappa = 0.289$. The corresponding HF-values yield
$< r^{-3} >$ = 7.737 and $\kappa = 0.455$.

Compared with the density functional {\it ab-initio} results for the 
embedded (CuO$_6)^{-10}$
ion, the single-hole picture already gives reasonable results for ${\rm a_{dip}}$ and
${\rm a_{cp}}$. It fails, however, in its prediction of the EFG with a
value of $V_{zz} = - 4/7 < r^{-3} >$.
This discrepancy has sometimes been used to assign a fractional hole number to the 
${\rm 3d_{x^2-y^2}}$ orbital. The underlying physical picture, however, is completely 
misleading since the
contributions to the EFG originate from various Cu shells as can be seen in 
Table~\ref{tab:contr}. The cluster EFG value of 1.4 as obtained with the DF method,
results from cancellations
between relatively large individual contributions. This shows that the theoretical
determination of EFGs is quite delicate. It is necessary to describe all electron
shells of the atom accurately.

At this point we would like to discuss the differences in the results as obtained in HF-
versus DF- theory by analyzing the details of the electronic structure.
For each spin projection there exists a total of 23 molecular orbitals (MO)
that can be formed as linear combinations of the Cu 3d and the six O 2p atomic
orbitals (AO). Both HF and DF predict that the MO with highest energy is the
anti-bonding hybridization between Cu 3d$_{\rm x^2-y^2}$ and O 2p$_{\rm x}$
and 2p$_{\rm y}$ AOs.
It is occupied for one spin projection (spin up) only and, in the following, energies
will be measured with respect to the energy of this highest occupied molecular
orbital (HOMO). (The HOMO calculated with DF for a cluster comprising 9 Cu atoms will
be shown in Fig.~\ref{homo9}.) 
The corresponding spin-down orbital is the lowest unoccupied molecular
orbital (LUMO) and lies at an energy which is at 1.4 eV for DF and even higher for HF. 
The energies of
all 23 MOs are shown in Fig.~\ref{mos} together with the contributions from the
individual AOs. The length of the bars marks the squared values of the expansion
coefficient of the MO with respect to the AOs whereby spin-up (spin-down) orbitals
are denoted by solid (dotted) bars.

The orbitals with second highest energy are anti-bonding linear combinations of
the AOs 3d$_{\rm 3z^2-r^2}$ of the Cu and O 2p$_{\rm z}$ of the apical oxygens.
In the DF calculation, these are followed at energies around -1.5 eV by MOs that
can be formed as anti-bonding hybridizations with the three other Cu 3d AOs.
Deeper in energy, between -2.6 and -4.4 eV, are the MOs that are composed of oxygens only
without contributions from Cu. They comprise the 2p$_{\rm z}$ AOs of the planar
oxygens and the 2p$_{\rm x}$ and 2p$_{\rm y}$ of the apex oxygens. 
At the HF level, these are at energies (-1.1 to -1.8 eV) that are above 
those of the MOs with contributions from Cu 3d$_{\rm xy,yz,zx}$.

The MO with lowest energy in Fig.~\ref{mos} is at -5.8 eV for DF. It is
formed by the bonding hybridization between 
3d$_{\rm x^2-y^2}$ and O 2p$_{\rm x}$ and 2p$_{\rm y}$.
Further down in energy lie 2 $\times$ 21 orbitals formed by inner-shell AOs.
Their energies and wave functions are only slightly different for the two
spin projections though it is precisely this difference that will determine
the core polarization and also contributes to the electric field gradient.
Therefore it is essential to describe also inner-shell electrons with
basis sets of sufficiently high quality.

The HF result suggests an almost purely ionic bonding with small overlap between Cu and O.
In contrast, the DF results emphasize the covalent character of the
bonds and an appreciable overlap. 
This varient description of bonding is also reflected in the width of the ``d-band''
(10 eV vs. 6 eV) and the localization of the atomic spin density at the Cu
(0.90 vs. 0.67).

The contributions of the DF MOs shown in Fig.~\ref{mos} to the density of states
is represented in Fig.~\ref{dos}.
The individual states were broadened by folding with a Gaussian function of
half-width 0.21 eV. A comparison to the density of states obtained from band structure
calculations~\cite{pickett,haas} shows an overall agreement even for the small cluster
under consideration.

The total occupation of the DF Cu atomic orbitals amounts to 8.92 
which results from the
following contributions: .08, .89, .96, .96, .92, .97 from 4s, 3d$_{\rm 3z^2-r^2}$,
3d$_{\rm zx}$, 3d$_{\rm xy}$, 3d$_{\rm x^2-y^2}$, 3d$_{\rm xy}$ with spin up and
.07, .66, .96, .96, .27, 1.00 with spin down. The small population of the Cu 4s AO is due
to the fact that the hybridization between the 3d$_{\rm 3z^2-r^2}$ and the 2p$_{\rm z}$
of the apex oxygens also involves a minor admixture of 4s orbitals. 
The populations obtained from the analysis of the HF results are somewhat higher, in
agreement with Sulaiman et al.~\cite{sulaiman}.

In conclusion, we think that the Cu-O bonding is better described by DF methods, 
which include part of the correlation, than with HF. This will gain further 
evidence from the results
obtained with larger clusters that will be reported in the following sections.
\section{Electric Field Gradients}
\label{efg}
\subsection{Larger clusters}
\label{largercl}
To investigate the convergence of the calculated EFGs and magnetic hyperfine 
properties with respect to cluster size, calculations for clusters comprising $n$
Cu atoms ($n = 1, 2, 4, 5, 9$) have been performed using the HF method as
well as DF 
with GGA (BLYP-functional). The multiplicity $m$ of the spin states was chosen 
to be maximal ($m = 2 n + 1$). Lower multiplicities were also investigated for
selected clusters ($ m = 0$ for $n = 2 $ and $ m = 2 , 4$ for $n = 5$). On all atoms
belonging to the cluster, a 6-311G basis set was employed. 
In Fig.~\ref{clustercu9} the largest cluster used is represented. It comprises 
9 Cu and 42 O atoms in the plane and is surrounded by 12 Cu and 50 La ions that
were treated with bare pseudopotentials. The corresponding HOMO as obtained
with DF methods is shown in Fig.~\ref{homo9}.

In this section, the 
resulting EFG values are discussed whereas the magnetic hyperfine properties will 
be analyzed in Sec.~\ref{hff}.
\subsection{Electric Field Gradients at the Cu Site}
\label{efgcu}
The results of Hartree-Fock and density-functional calculations for the
V$_{zz}$ component of the EFG at the Cu site are given in Table~\ref{tab:cuefg}.
Since the Cu sites in the two clusters with even number of Cu atoms are not situated
in the center, a small asymmetry $\eta$ results.
Again, the large discrepancy between HF and DF results is obvious.
Our HF values for small clusters are in agreement with those obtained in 
Refs.~\cite{sulaiman,martin}.

The variations of the values for the EEGs with respect to cluster size and
multiplicity are within reasonable bounds considering the subtle cancellations
of contributions from the various shells as can be seen in Table~\ref{tab:contr}.

The experimental value~\cite{imaisl} for the $^{63}$Cu quadrupole frequency is 
33.0 MHz. This corresponds to a value for $V_{zz}$ of 1.331 a.u. using 
a quadrupole moment of $^{63}Q = -0.211$ b,
but to 1.560 a.u. with $^{63}Q = -0.18 $b. The former value for $^{63}Q$ is from an
analysis of Sternheimer~\cite{sternheimer} and is the one cited in the current
NQR tables while the latter was determined by Stein et al.~\cite{stein} from
a HF cluster calculation for cuprite. In relation to the large amount of NQR data
available on cuprate superconductors, the availability of reliable values for
the quadrupole moments of Cu would be very desirable.

In a comparison with experiments it should further be noted that our calculations
were performed for clusters with atomic positions corresponding to the 
tetragonal phase. Moreover, spin states with lower multiplicities better account
for antiferromagnetic fluctuations than high-spin states. In this respect, our
calculated values are by 17 \% too low, even with $^{63}Q = -0.211 b$. 
Nevertheless, it can be concluded that a valuable estimate for $V_{zz}$ can
already be gained with DF calculations with the cluster CuO$_6$/Cu$_4$La$_{10}$. 
This is of
relevance for studies of the changes in the quadruole frequency upon doping.

\subsection{Electric Field Gradients at the O Site}
\label{efgo}
Although the clusters used in this report were constructed to
investigate mainly the magnetic hyperfine interactions at the Cu site, the results
obtained for the EFGs at the oxygen sites  are also of interest. For the planar 
oxygen O(p) they are listed in Table~\ref{tab:opefg}. Again, we find
reasonable agreement between the experimental values and our calculations 
with the DF method.

For  O(a), the apical oxygen, the EFG is axially symmetric and the
component $V_{zz}$ turned out to have the
values .26, .21, and .19 for DF (.28, .26, and .19 for HF) for 
clusters with 4, 5, and 9 Cu atoms, respectively.
The experimental value~\cite{ishida} is .22.

\section{Core-polarization and transferred hyperfine fields}
\label{hff}
\subsection{Isotropic Contributions}
The magnetic hyperfine interaction at a nucleus can be decomposed into an isotropic
part D and a traceless dipolar part. The former is given by the difference
between the spin densities at the nuclear site R:
\begin{equation}
{\rm D(R) = \frac{8 \pi}{3} \left( \sum_m \mid \psi_m^{\uparrow}(R) \mid^2
                                - \sum_{m'} \mid \psi_{m'}^{\downarrow}(R) \mid^2 \right)} 
\end{equation}
where the sums extend over all occupied MOs.
This quantity is called Fermi contact contribution if the singly occupied MOs contain
AOs of s-symmetry centered at R.
It is called
core-polarization if the MOs are doubly occupied but with differing
contributions from the s-like AOs. In the present case, the highest
occupied molecular orbital is always composed of a hybridization of predominantly
atomic Cu 3d$_{x^2-y^2}$ and O 2p$_{x,y}$ character. Therefore we prefer 
the use of the notation core-polarization in the following.

The values for D(Cu) vary strongly with cluster size and position of a particular Cu
atom in larger clusters. In the Cu$_9$O$_{42}$/Cu$_{12}$La$_{40}$ cluster, e.g.,
D = 0.922  for the central Cu which is surrounded by 4 nearest neighbors (NN),
but D =  $-0.364$ at the four equivalent corner sites which have 2 NN and D = 0.321
at the four sites on the edges with 3 NN. The complete set of data is 
compiled in Table~\ref{asd}.
In Fig.~\ref{Dvsn}, the values for D(Cu$_i$) obtained with clusters of different sizes
are plotted vs. N, the number of nearest neighbor Cu atoms of Cu$_i$.  
In all calculations, the same basis set 6-311G was used.
It is obvious that in a first approximation the increase of D(Cu) is linear
in the number of nearest neighbor Cu atoms for both DF and HF methods.
It is suggestive to attribute this change to a transferred hyperfine term b and to
decompose D(Cu) into 
\begin{equation}
{\rm D(Cu)  = a +  b \; N}
\end{equation}
with preliminary values: a $\approx -1.75$ and b $\approx$ 0.69 for DF/BLYP and 
     a $\approx -3.52$ and b~$\approx~0.17$ for HF.

For those oxygen sites that are at the border of the cluster and that are therefore
bounded to a single Cu atom, a value D(O) $\approx 0.64$ is found. For oxygens
bridging two Cu atoms, however, we get  D(O) $\approx 1.24$ with 
the HF values 0.60 and 1.11 being only slightly lower.
Therefore, also the core polarization at O is determined by a transferred hyperfine
field from its neighboring Cu atoms:
\begin{equation}
{\rm D(O)  =  c \;  N}  
\end{equation}
with a preliminary value $c \approx 0.6$.

At the apex oxygens, the spin density is marginal (on the order of 1 \%) and the
core polarization amounts to 0.05.

In retrospective, some of the early problems in analyzing NQR and NMR data on
cuprate superconductors can be traced back to the fact that the single-hole model
developed in Ref.~\cite{Bleaney} for an isolated Cu$^{2+}$ ion in a crystal field, 
cannot be applied to correlated Cu ions in a CuO$_2$ plane. To be precise, it works reasonably well
for local quantities like the dipolar coupling tensor and the spin-orbit interaction
which will be discussed later. Their values are mainly determined by the HOMO and 
also do not depend too sensitively on the method of treating the many-electron system.
Even HF calculations give results of the correct order of magnitude. The core-polarization,
however, is not a local affair. As Fig.~\ref{Dvsn} exhibits, hyperfine fields
transferred from the neighboring copper ions have a drastic effect on D(Cu). Its value
is negative for an isolated copper ion but is increased by contributions from
neighboring ions. At the HF level, there is a small hybrization and the transfer
is small. DF calculations, however, render a stronger covalency and therefore 
the spin density is more delocalized which implies
a larger transfer. With contributions from four NN Cu ions, D(Cu) becomes
positive as was first recognized by Mila and Rice~\cite{milarice}.

The question then arises whether also contributions from further distant Cu atoms
will contribute to the hyperfine properties at Cu or O. This is of particular relevance
since to reconcile NMR and neutron-scattering experiments within a one-component
spin model, Zha et al.~\cite{zha} have advocated a hyperfine field at oxygen
transferred from second nearest Cu neighbors. This would lead to a reduction of
the form factor $^{17}F(\vec{q})$ around the wave-vector $\vec{q} = ( \pi /a, \pi /a)$
and to a modification of the spin-lattice relaxation rate $^{17}T_1^{-1}$.

To investigate this problem of hyperfine fields transferred from further distant
neighbors in detail, we performed also DF cluster calculations for spin states
with lower multiplicity. In the cluster Cu$_5$O$_{26}$/Cu$_8$La$_{34}$ 
states with multiplicities $m = 4$ and $m = 2$ were obtained. 
For the former the spin density at the central Cu atom is of opposite sign to that 
of the four neighboring sites. For $m = 2$, it is negative at the central and one 
edge Cu atom. 
In Fig.~\ref{spind} the corresponding differences in the spin densities along the
Cu-O-Cu-O-Cu axes are shown. They peak at positions where the squares of the 
3d$_{x^2-y^2}$ orbitals have their maxima. The core-polarization then yields a
change close to the Cu nucleus. In the ``antiferromagnetic'' cluster that is 
obtained with m = 4, the atomic spin density at the central Cu is negative but 
its magnitude is smaller than those at the four adjacent coppers (the detailed results
are given in Table~\ref{asdm}). This is due to the finite size of the cluster.
There is an excess of positive spin density in the central region which is
also reflected by the atomic spin densities at the four oxygens that are
bridging two antiferromagnetically coupled Cu atoms. There, a value of 0.026 is
obtained instead of the cancellation expected for a large system.

To account for these finite size effects, we have carefully analyzed all results
given in Tables~\ref{asd} and ~\ref{asdm}. It turned out that the values of the
core-polarization at the oxygens are proportional to the sum of atomic
spin densities at the adjacent Cu ions
\begin{equation}
{\rm	D(O) = \gamma \sum_{j \in NN} \rho(Cu_j) }
\end{equation}
where the sum extends over one or two NN. This is illustrated in Fig.~\ref{DO}.
The negative values result from the spin states with lower multiplicities. The
deviations from the linear behavior for some small D(O)-values is due to errors
in forming the difference between large ${\rm \rho(Cu_j)}$ values with opposite signs.

Similarly, the core-polarization at the Cu ions can well be reproduced by decomposing
it according to
\begin{equation}
{\rm	D(Cu_i) = \alpha \rho(Cu_i) + \beta \sum_{j \in NN} \rho(Cu_j)\quad . }
\end{equation}
In Fig.~\ref{Dcu} the various values ${\rm Du(Cu_i)/\rho(Cu(i))}$ are plotted
as a function of the sum of NN atomic spin densities divided by ${\rm \rho(Cu_i)}$.

With all these data available, we are now in a position to study the intrinsic
and transferred hyperfine fields in a quantitative manner and to estimate also the
influence of next-nearest and further distant Cu atoms. 
The isotropic hyperfine constants at the oxygens were assumed to depend on the
spin densities of the adjacent (NN), next-nearest-neighbors (NNN) 
and further distant Cu atoms according to

\begin{equation}
{\rm D(O) = \gamma \sum_{j \in NN} \rho(Cu_j) + \gamma' \sum_{j \in NNN} \rho(Cu_j) 
		+ \gamma'' \sum_{j \in NNNN} \rho(Cu_j) .}
\label{eqDO}
\end{equation}
For Cu, the isotropic hyperfine constants were decomposed into

\begin{equation}
{\rm D(Cu_i) = \alpha \rho(Cu_i) + \beta \sum_{j \in NN} \rho(Cu_j) + 
	\beta' \sum_{j \in NNN} \rho(Cu_j) + \beta'' \sum_{j \in NNNN} \rho(Cu_j)} .
\label{eqDCU}
\end{equation}

Figs.~\ref{DO} and \ref{Dcu} already show that the main contributions are due to NN.
This result is corroborated by least squares fits of all data to
Eqs.~(\ref{eqDO}) and (\ref{eqDCU}) which yield

\begin{equation}
	\alpha = -2.54 \pm 0.04 \qquad \beta = 1.02 \pm 0.02 
	\qquad \mid \beta' \mid \; \le 0.02 \qquad \mid \beta'' \mid \; \le 0.02 
\end{equation}
and
\begin{equation}
	\gamma = 0.925 \pm 0.006 \qquad \mid \gamma' \mid \; \le 0.006
		\qquad \mid \gamma'' \mid \; \le 0.006 \quad.
\end{equation}
As expected, the contributions from Cu ions beyond NN are marginal.

The analyses performed so far have demonstrated that all hyperfine coupling parameters are
proportional to the sum of the atomic spin densities at the adjacent Cu ions. The essential
point is that in this way also the values calculated for lower spin multiplicities could
be accounted for. These contain complementary informations to the results of high-spin
states which strongly improve the quality of the results as can be seen 
in Figs.~\ref{DO} and \ref{Dcu}. Incorporating negative values of ${\rm \sum_j \rho(Cu_j)}$
into the evaluation renders a much better understanding of the various contributions.

We are now in a position to extrapolate the data from the finite clusters to an
extended system.
Considering the three different values of ${\rm \rho(Cu_i)}$ in the largest 
cluster (see Table~\ref{asd}), we expect ${\rm \rho(Cu)}$ in the homogeneous case to be 0.70.

This leads to the following values:
a = $-1.778$,
b = 0.714,
$\mid b' \mid \le 0.014 $, 
c = 0.648,
and
$\mid c' \mid \le 0.004$.

It should be emphasized again, that the extracted values for the contact terms 
depend on the particular choice of the exchange-correlation functional. We have performed a few
calculations with other functionals, too. They deliver values that differ by 15 \% but
they also lead to the conclusion that contributions from further distant neighbors are
not important.
\subsection{Dipolar Hyperfine Interactions}
In contrast to the core-polarization, which is proportional to the spin density
difference at a single point, the dipolar
hyperfine coupling results from a spatial average of $1/r^3$ with wave-functions.
Its value, therefore, can be determined with much greater accuray than that of the contact
interaction and does less crucially depend on the calculational method or the choice of the
exchange-correlation functionals. 
It is given by
\begin{equation}
{\rm a^{ij}_{dip}(R) = \sum_{m}   \left< \psi_m^{\uparrow}(r) \mid T^{ij}(r-R)
		 \mid \psi_m^{\uparrow}(r) \right> - \sum_{m'} \left< 
	\psi_{m'}^{\downarrow}(r) \mid T^{ij}(r-R) \mid \psi_{m'}^{\downarrow}(r) \right> }
\end{equation}
where
\begin{equation}
{\rm	T^{ij}(x) = \frac{3x_i x_j - \delta_{ij} x^2}{x^5} \; .}
\end{equation}
In Table~\ref{tab:adip} the values  for ${\rm a^{\parallel}_{dip}}$ (for Cu)
and  ${\rm c^{\parallel}_{dip}}$ (for O) obtained with various clusters are given. 
It should be remarked that 
${\rm c_{dip}^{\parallel}}$ for planar oxygen refers to the directions along the Cu-O-Cu
bonds. The tensor coupling is only nearly axially symmetric.

In analogy to the isotropic case, the dipolar contributions also turned out to depend 
mainly on the atomic spin densities of the NN copper ions besides the on-site term for
Cu which is proportional to the on-site atomic spin density. A least squares fit of
all data (including those with lower multiplicities) to the ansatz

\begin{equation}
{\rm a_{dip}^{\parallel} (Cu_i) = \alpha_{dip} \rho(Cu_i) + 
	\beta_{dip} \sum_{j \in NN} \rho(Cu_j)   +
	\beta'_{dip} \sum_{j \in NNN} \rho(Cu_j) + 
	\beta''_{dip} \sum_{j \in NNNN} \rho(Cu_j)} 
\end{equation}
gives ${\rm \alpha_{dip} = -5.206 \pm 0.014, \beta_{dip} = 0.101 \pm 0.006}$, and
${\rm \mid \beta'_{dip} \mid \le 0.005}$. Thus there is also a small transferred dipolar
hyperfine interaction for Cu.

For the oxygens, the ansatz 

\begin{equation}
{\rm c_{dip}^{\parallel}(O) = 
	\gamma_{dip} \sum_{j \in NN} \rho(Cu_j)   +
	\gamma'_{dip} \sum_{j \in NNN} \rho(Cu_j) + 
	\gamma''_{dip} \sum_{j \in NNNN} \rho(Cu_j)} .
\end{equation}
leads to
${\rm \gamma_{dip} = 0.527 \pm 0.006, \mid \gamma'_{dip} \mid \le 0.006}$, and
${\rm \mid \gamma''_{dip} \mid \le 0.006}$. 

Performing the same extrapolation as was applied to the isotropic contributions, we finally get
${\rm a_{dip}^{\parallel} = -3.644}$,
${\rm b_{dip}^{\parallel} = 0.071}$,
${\rm c_{dip}^{xx} = 0.369}$ (along the bond direction),
${\rm c_{dip}^{yy} = -0.177}$ (perpendicular to the bond),
and
${\rm c_{dip}^{zz} = -0.192}$ (perpendicular to the plane).

Represented in Table~\ref{tab:compilo} are the magnetic hyperfine couplings 
calculated for O(p).
The spin-orbit contributions to the hyperfine fields are expected to be small
in the case of O. We neglect them completely by assuming 
${\rm c^{ii}_{tot} = c + c^{ii}_{dip}}$. These values may be compared to those which have
been extracted from various experiments in Refs.~\cite{zha} and~\cite{russ1}.
\subsection{Origin of the Transferred Hyperfine Fields}
\label{origin}
To investigate the mechanisms of spin transfer, we will analyze in detail the results from the two
clusters containing one and two copper atoms and point out the differences.
We commence by noting that the basic notions of spin-polarized density functional theory~\cite{kohn} 
rely on the concept of expressing all physical relevant quantities in terms of the
spin densities alone. The ``wave functions'' are just auxiliary entities that are
introduced to solve the Kohn-Sham equations~\cite{kohnsham} and have no direct
physical meaning. Nevertheless, to understand the origins of transferred hyperfine
fields, it is convenient to assign the AOs which are built up from localized basis
functions to atomic single-electron wave functions. Similarly, the MOs are interpreted
as hybridizations between the individual AOs. Another remark concerns the representation
of the AOs in terms of basis functions. We recall that the employed basis sets describe
a ``single-electron wave function'' as a linear combination of several contracted radial
Gauss functions. For Cu, there are three radial functions for each of the five 3d-orbitals
and nine functions of s-like character. For O, we have four s-functions and three radial functions
for each of the 2p-orbitals.

Let us first consider the (CuO$_6$)$^{10-}$ ion embedded in the appropriate lattice. 
Overlap and covalent effects convey spin density from the
Cu$^{2+}$ ion onto the ligand oxygen sites whose spin direction is parallel to that of the local
Cu moment. The data in Table~\ref{cuo6} show that in the DF calculations 
the atomic spin density $\rm \rho(Cu)$ is reduced from 1 to 0.667 in favor of a spin density transfer
to the four planar oxygens 
($4 \times {\rm \rho(O(p))}$ = 0.328). The transfer from the Cu to the O(p) is much
less pronounced at the HF level where ${\rm \rho(Cu)}$ = 0.902. 
This trend is also reflected in the dipolar hyperfine coupling (see Table~\ref{tab:adip}):
${\rm a_{dip}^{xx} = 1.763~(2.210)}$ and ${\rm c_{dip}^{xx} = 0.388~(0.181)}$ for DF (HF)
with a total of  ${\rm a_{dip}^{xx} + 2 \times c_{dip}^{xx} = 2.539~(2.572)}$. The transferred
spin density is mainly on the O 2p$_{\sigma}$ orbital, but a small amount also goes to the O 2s which
is thus expected to be polarized parallel to the spin density on the Cu. Indeed, we
get ${\rm D_{2s}(O) = 0.772}$, as can be seen
in Table~\ref{tab:fermio} where the core polarizations ${\rm D(O)}$ from DF calculations
are listed for the individual orbitals. The 2s orbital in turn
polarizes the 1s but with opposite sign (${\rm D_{1s}(O) = -.141}$). 
Since the MOs are linear combinations of AOs, the contact term may also
have contributions that come from a product of an s-like AO centered at the nucleus 
under consideration with an AO centered on an adjacent atom. These contributions turn out
to be small. They are listed separately as numbers in parentheses
in Tables~\ref{tab:fermio} and~\ref{tab:fermicu}. On the Cu, the delocalization of 
spin densities implies the following changes.
The MO second highest in energy (at $-.14$ eV) is occupied with a spin-down electron.
It mainly mixes the Cu 3d${\rm _{3z^2-r^2}}$ with the O(a) 2p$_z$ but has also some small
contributions from the Cu 4s and the O(p) 2p$_{\sigma}$ orbitals.
The admixture of the 4s is larger than that
of the following orbital (at $-.58$ eV) which has the same symmetry but is occupied with a spin-up
electron. These two MOs thus cause the negative value 
${\rm D_{4s}(Cu) = -.650}$ (see Table~\ref{tab:fermicu}).
The 4s AO in turn polarizes the inner s orbitals. The alternating signs of the 
4s, 3s, and 2s contributions~\cite{engels88} are a consequence of the Pauli exclusion principle.

We next consider the system consisting of 2 Cu atoms with parallel moments
and with 1 bridging O (O$_{\rm br}$)
and 6 planar O (O$_1$) bonded to one of the Cu only. The DF calculations yield the atomic spin
densities ${\rm \rho(Cu)}$ = 0.668, ${\rm \rho(O_{br})}$ = 0.130, and ${\rm \rho(O_1)}$ = 0.089.
Since the 2p$_{\sigma}$ orbital can only share a maximal amount of hole contribution,
the additional transfer of spin density onto the O$_{\rm br}$ is less than twice the value
in the CuO$_6$ system. The missing fraction is added to the six O$_1$.  
More positive spin density, however, can
be put on the 2s which leads to the total value ${\rm D(O_{br}) = 1.275}$ (see 
Table~\ref{tab:fermio}). The dipolar contributions are
${\rm a_{dip}^{xx} = 1.746}$ and ${\rm c_{dip}^{xx} = 0.324}$.
On the Cu atoms, the spin density distributions in the inner s shells cannot be changed
significantly compared to the CuO$_6$ case since not much redistribution is possible in the
inner core. Therefore the additional positive spin density resides mainly on the 4s atomic orbital.

We thus get a coherent picture of the spin density distributions for a single Cu$^{2+}$ ion
surrounded by four planar oxygens and of the additional spin transfers caused by a second copper
ion. The interpretation in terms of localized AOs turned out to be helpful for the
understanding of the spin densities which are the relevant quantities in density functional
theory. The results of our quantum chemical calculations demonstrate that the spin densities, 
when attributed to the Cu or to the O, are strongly connected.
In the ground state, there is no reason for advocating a two-component model and we also
see little chance that low-energy excited states could change this fact.
\subsection{Spin-Orbit Coupling}
For Cu, the spin-orbit coupling gives an
appreciable contribution to the total hyperfine fields, especially in the perpendicular
direction since accidentally a and ${\rm a^{\perp}_{dip}}$ almost cancel 
(${\rm a + a^{\perp}_{dip} = 0.003}$). 
A calculation of the spin-orbit interaction cannot be carried out at the same level
of quality as is possible for the other hyperfine interactions.
Therefore we adopt the values for the isolated Cu$^{2+}$ ion from Ref.~\cite{Bleaney} in
the simplified form used in Ref.~\cite{mps}:
\begin{equation}
{\rm	a^{\parallel}_{so} = - \frac{62}{7} k < r^{-3} > }
\end{equation}
and
\begin{equation}
{\rm	a^{\perp}_{so} = - \frac{11}{7} k < r^{-3} > }
\end{equation}
with a parameter value k = $-0.044$. With ${\rm < r^{-3} > = 6.171}$, we get 
${\rm a^{\parallel}_{so}}$ = 2.405 and 
${\rm a^{\perp}_{so}}$ = 0.427.
It should be noted that the uncertainty in the spin-orbit coupling parameter and the
excitation energies is comparatively large. Therefore, in Ref.~\cite{mps} a range
of values for k of $\pm$ 20 \% was considered. 

In Table~\ref{tab:compilcu} all our calculated values for the hyperfine coupling parameters 
for Cu are compiled. The spin-orbit contributions are put in parentheses to emphasize that
they are estimated.

No direct experimental information is available on these parameters. They can be determined
indirectly from a combination of anisotropic Knight shifts and relaxation rates and the
nuclear resonance frequencies in antiferromagnetic compounds. The results of various
analyses are shown in Table~\ref{tab:compilcu}. Our values are in reasonable agreement with the data. 
\section{Summary and Conclusions}
\label{con} 
The electronic structure of La$_2$CuO$_4$ has been investigated by first-principles
cluster calculations. These have the advantage that local properties can be studied
in great detail but usually suffer from a somewhat uncontrolled embedding in the
periodic lattice and surface problems. By using a sequence of clusters containing
up to nine copper atoms, we have demonstrated which quantities can be evaluated with clusters 
of modest size.

As concerns the theoretical approximations to treat the many-electron problem in
CuO$_2$ planes, we have shown that distinct differences between the Hartree-Fock
and density functional methods exist. From our results we conclude that the former approach,
which neglects correlations entirely, is not adequate for cuprate superconducting materials.

We have evaluated the electric-field gradients at the Cu and the O sites for a variety
of cluster sizes and spin multiplicities. For both planar and apical oxygens a satisfactory
agreement with the experimentally determined quadrupole frequencies was obtained. For copper,
a comparison between theoretical and experimental values is hampered by the uncertainty
with which the Cu nuclear quadrupole moments are known.

With respect to magnetic hyperfine properties one has to distinguish between the 
core-polarization and the dipolar contributions. The latter can be determined theoretically
with good reliability. The former, however, is a quantity whose value is
given by subtle cancellations of contributions from various s-like atomic orbitals.
Our calculations demonstrated that besides the negative on-site core-polarization for Cu,
a sizeable positive contribution transferred from neighboring Cu$^{2+}$ ions exists as
has been suggested by Mila and Rice~\cite{milarice}. 

We have investigated this transfer of spin density in detail and found 
that there is no appreciable contribution from copper ions other than the 
nearest neighbors. This is the main result of the present
work since it has far reaching consequences. In particular, it questions the reconciliation 
between NMR and neutron scattering experiments and points out that the
disagreements are now as before. By introducing a transferred hyperfine coupling
between next-nearest neighbor Cu spins and $^{17}$O nuclei spin, 
Zha et al.~\cite{zha} were able to conciliate many different experiments.
We have shown that there is no microscopic justification for the presence of these
additional hyperfine couplings. The discrepancies among the data therefore
need another explanation.

On the other hand, the results of the analysis of the origins of the calculated hyperfine
fields carried out in Sec.~\ref{origin} shows that the spin densities on the coppers
and oxygens are very tightly connected. This gives strong support for a one-component
model of the spin fluid, at least as concerns ground-state properties. 

We have reached these conclusions by carrying out extended ab-initio cluster calculations.
One has, of course, to take into consideration that this approach to describe the electronic
structure emphasizes local features and is surely far more appropriate for insulating materials
than for metals. It would be a big surprise, however, when a tight-binding picture
completely failed to reproduce qualitative aspects of a system with itinerant charge carriers.
This would shift the problems to a quite different field.

The calculated values for the magnetic hyperfine couplings are in agreement with those 
extracted from various NMR and NQR experiments if we adopt the conventional estimates
for the spin-orbit interaction for the Cu$^{2+}$ ion. In this respect, improved quantum
chemical calculations of the spin-orbit coupling would be desirable. Although our results 
give a small value for the hyperfine field that contributes to the spin part of the Knight
shift in c-direction (${\rm a^{\parallel} + 4 b + 4 b^{\parallel}}$), we think that
the vanishing of ${\rm K_s^{\parallel}}$ below $T_c$ in all high-temperature superconductors
and with all doping concentrations is still another open problem that deserves further
experimental and theoretical consideration.
\acknowledgments
We would like to thank D. Brinkmann, M. Mali, S. Schafroth, and J. M. Singer for
helpful comments and critical reading of the manuscript. One of us (P.F.M.) would 
like to thank C. P. Slichter for enlightening discussions.
This work was partially supported by the Swiss National Science Foundation. 
A major part of the computation was carried out at the national supercomputing center CSCS.


\begin{figure}
\caption {
The CuO$_6$/Cu$_4$La$_{10}$ cluster. 
}
\label{clustercu1}
\end{figure}

\begin{figure}
\caption {
Energies of the highest occupied MOs in the CuO$_6$/Cu$_4$La$_{10}$ cluster
with contributions from the individual AOs: (a) HF, (b) DF. The length of the bar is proportional
to the square of the expansion coefficient of the MO into the corresponding AOs.
Spin-up (spin-down) orbitals are denoted by solid (dashed) bars.
}
\label{mos}
\end{figure}

\begin{figure}
\caption {Total and partial density of states of the highest ``band'' in 
the CuO$_6$/Cu$_4$La$_{10}$ cluster as obtained with the DF method.  
}
\label{dos}
\end{figure}

\begin{figure}
\caption {
The Cu$_9$O$_{42}$/Cu$_{12}$La$_{50}$ cluster. 
}
\label{clustercu9}
\end{figure}

\begin{figure}
\caption {
Highest occupied molecular orbital for the 
Cu$_9$O$_{42}$/Cu$_{12}$La$_{50}$ cluster. 
}
\label{homo9}
\end{figure}

\begin{figure}
\caption {
Difference between spin-up and -down densities at Cu vs. number of nearest neighbor
Cu atoms, $N$, for different sites in various clusters with $n$ Cu atoms.
Circles: $n = 1$, diamonds: $n = 3$, triangle down: $n = 4$, triangle up: $n = 5$, 
squares: $n = 9$. DF (HF) values correspond to the upper (lower) data set.
}
\label{Dvsn}
\end{figure}

\begin{figure}
\caption {
Difference in spin densities along the 
O-Cu-O-Cu-O-Cu-O bonds as calculated
with DF/GGA for the cluster Cu$_5$O$_{26}$/Cu$_8$La$_{34}$. From top to bottom: 
multiplicity $m = 6$, $m = 4$, and $m = 2$ horizontally (vertically) with spin
density signs $+ - + \quad ( - - + ).$
}
\label{spind}
\end{figure}

\begin{figure}
\caption {
Difference between spin-up and -down densities at O(p) vs. sum of atomic
spin densities at neigboring Cu atoms for different sites in various 
clusters: Cu$_9$ (diamonds), Cu$_5$, m=6 and m=4 (circles), Cu$_4$ (squares), and
Cu$_5$ with m=2 (open circles).
}
\label{DO}
\end{figure}

\begin{figure}
\caption {
$\rm D(Cu_i)/\rho(Cu_i)$ values plotted against the sum of atomic spin
densities at the NN Cu atoms divided by $\rm \rho(Cu_i)$:
Cu$_9$ (diamonds), Cu$_5$, m=6 and m=4 (circles), and
Cu$_5$ with m=2 (open circles).
}
\label{Dcu}
\end{figure}

\begin{table}[h,b] \centering
\caption[dummy]{
\baselineskip=24pt
Clusters, number of atoms N, atoms with pseudopotentials PP, total number of 
electrons E,
number of basis functions B, and number of primitive Gaussian functions P.
}
\vspace{5mm}
 \begin{tabular}{lrrrrr}
 cluster        & N  &   PP \qquad         & E $\;$  & B \quad & P \quad \\ 
 \hline
 CuO$_6$        &  7 &   Cu$_4$La$_{10}$  &  87 & 117 & 222 \\
 Cu$_2$O$_{11}$ & 13 &   Cu$_6$La$_{12}$  & 164 & 221 & 418 \\
 Cu$_4$O$_{20}$ & 24 &   Cu$_8$La$_{26}$  & 308 & 334 & 772  \\
 Cu$_5$O$_{26}$ & 31 &   Cu$_8$La$_{34}$  & 395 & 533 & 1006 \\
 Cu$_9$O$_{42}$ & 51 & Cu$_{12}$La$_{50}$ & 663 & 621 & 1116 \\
 \end{tabular}
\label{clust}
\end{table}

\begin{table}
 \caption[dummy]{
\baselineskip=24pt
EFG component V$_{zz}$, 
core polarization ${\rm a_{cp}}$, 
dipolar contribution to the hyperfine tensor a${\rm _{dip}^{\parallel}}$, and 
atomic spin density $\rho$ at the Cu, as obtained with HF and 
DF/GGA calculations for the cluster CuO$_6$/Cu$_4$La$_{10}$.}
\vspace{5mm}
 \begin{tabular}{llrrrc}
 method        & \quad basis &  $V_{zz}$ \, &  $a_{cp}$ \, & 
 a$_{dip}^{\parallel}$ \, & $\rho$ \\
 \hline
 HF $^a$       &          & 2.232  & -0.503 & -4.421 & .88 \\
 HF $^b$       &          & 1.840  &        &        &      \\\hline
 HF            & 6-311G   & 1.936  & -3.519 & -4.421 & .902    \\
 DF/BLYP       & 6-311G   & 1.419  & -1.784 & -3.526 & .667    \\ \hline
 DF/BLYP       & 6-311Gpd & 1.360  & -1.784 & -3.502 & .661  \\
 DF/SVWN       & 6-311G   & 1.364  & -1.625 & -3.425 & .642  \\
 DF/XALYP      & 6-311G   & 1.367  & -1.801 & -3.409 & .635  \\
 DF/BPW91      & 6-311G   & 1.415  & -1.943 & -3.517 & .669  \\
\end{tabular}
\label{cuo6}
$^a$ Ref. \onlinecite{sulaiman}\\
$^b$ Ref. \onlinecite{martin}\\
\end{table}

\begin{table}
\caption{ Contributions to the EFG component $V_{zz}$, calculated with
DF/GGA for the cluster CuO$_6$/Cu$_4$La$_{10}$,
from the nuclei within the cluster, from the point charges around the cluster, and
from the individual shells. The ``remainder''
lists just the (small) rest of all shells not explicitly given in the table. 
  }
\begin{tabular}{lrrrr}
Nuclei                       &            &             &   0.375    \\
Point charges                &            &             &   0.012    \\ \hline
                             & spin up &  spin down & sum \qquad \\ \hline
p$_x$, p$_y$                 &  $-1.183$ &  $-1.210$  &  $-2.393$  \\
p$_z$                        &   0.384   &    0.386   &    0.770   \\
d$_{x^{2}-y^{2}}$, d$_{xy}$  &  $-9.443$ &  $-5.720$  & $-15.163$  \\
d$_{3z^{2}-r^{2}}$           &   4.446   &    4.381   &    8.827   \\
d$_{xz}$, d$_{yz}$           &   4.522   &    4.481   &    9.003   \\
Remainder                    &   0.027   &  $-0.039$  &  $-0.012$  \\ \hline
Total                        &            &             &    1.419   \\
\end{tabular}
\label{tab:contr}
\end{table}

\begin{table}
\caption{
EFG component V$_{zz}$ and asymmetry parameter $\eta$ for
the Cu site in the La$_2$CuO$_4$ system calculated for different cluster sizes
and various spin multiplicities m. The basis set used for all calculations
was 6-311G. For comparison, the theoretical results of Sulaiman et al.~\cite{sulaiman} 
and Martin~\cite{martin},
as well as the measured quadrupole frequency~\cite{imai}, have been transformed
into atomic units.
}
\begin{tabular}{cccccc}
 & & \multicolumn{2}{c}{HF}  &\multicolumn{2}{c}{LDA/GGA} \\
cluster                         & m & V$_{zz}$ & $\eta$ & V$_{zz}$  & $\eta$ \\ \hline
CuO$_{6}$/Cu$_{4}$La$_{10}$         & 2 & 1.936   &   0    &   1.419   &    0   \\
Cu$_{2}$O$_{11}$/Cu$_{6}$La$_{12}$  & 3 & 1.904   &  0.04  &   1.360   &   0.14 \\
Cu$_{2}$O$_{11}$/Cu$_{6}$La$_{12}$  & 1 & 1.884   &  0.04  &   1.181   &   0.14 \\
Cu$_{4}$O$_{20}$/Cu$_{8}$La$_{26}$  & 5 & 1.925   &  0.00  &   1.354   &   0.01 \\
Cu$_{5}$O$_{26}$/Cu$_{8}$La$_{34}$  & 6 & 1.963   &   0    &   1.128   &    0   \\ 
Cu$_{5}$O$_{26}$/Cu$_{8}$La$_{34}$  & 4 &         &        &   1.071   &    0   \\ 
Cu$_{5}$O$_{26}$/Cu$_{8}$La$_{34}$  & 2 &         &        &   1.097   &   0.28 \\ 
Cu$_{9}$O$_{42}$/Cu$_{12}$La$_{40}$ &10 & 1.975   &   0    &   1.264   &    0   \\ \hline
CuO$_{6}^{a}$ & & 2.234 & 0.02 & \multicolumn{2}{c}{}  \\
CuO$_{6}$/Cu$_{4}$La$_{10}^{b}$ & & 1.840 & \multicolumn{3}{c}{} \\
Cu$_{2}$O$_{11}$/Cu$_{6}$La$_{12}^{b}$ & & 1.816 & \multicolumn{3}{c}{}  \\ \hline
experiment$^{c}$ &  &  \multicolumn{2}{c}{\qquad 1.560 ($^{63}$Q = $-0.18$ b)}& 
		\multicolumn{2}{c}{\qquad 1.331 ($^{63}$Q = $-0.211$ b)} \\
\end{tabular}
\label{tab:cuefg}
$^a$ Ref. \onlinecite{sulaiman}\\
$^b$ Ref. \onlinecite{martin}\\
$^c$ Ref. \onlinecite{imaisl}
\end{table}

\begin{table}
\caption{
EFG components for the planar oxygen O(p).}
\begin{tabular}{crrrrrr}
    &\multicolumn{3}{c}{HF}  &\multicolumn{3}{c}{LDA/GGA} \\
cluster                              & V$_{xx}$ & V$_{yy}$&V$_{zz}$& V$_{xx}$ & V$_{yy}$&V$_{zz}$\\ \hline
Cu$_{2}$O$_{11}$/Cu$_{6}$La$_{12}$ & $-.416$ & .335 & .081 & $-.858$  & .518  & .340 \\
Cu$_{4}$O$_{20}$/Cu$_{8}$La$_{26}$ & $-.403$ & .363 & .040 & $-.839$  & .536  & .303 \\
Cu$_{5}$O$_{26}$/Cu$_{8}$La$_{34}$ & $-.411$ & .326 & .049 & $-.847$  & .545  & .302 \\
Cu$_{9}$O$_{42}$/Cu$_{12}$La$_{50}$& $-.499$ & .353 & .146 & $-.881$  & .554  & .327 \\ \hline
experiment$^{a}$  & $-.75$  & $.51$  & $.24$   & $-.75$  & $.51$  & $.24$ \\
\end{tabular}
\label{tab:opefg}
$^a$ Ref. \onlinecite{ishida}\\
\end{table}

\begin{table}
\caption{Atomic spin densities and D-values at different planar sites in the
cluster Cu$_{9}$O$_{42}$/Cu$_{12}$La$_{50}$ as obtained with the DF method.
N denotes the number of nearest neighbor Cu ions.}
\begin{tabular}{crrr}
position   & N  & $\rho$ &  D \qquad \\ \hline
Cu(0/0)    & 4  & .6935  & 0.9215 \\
Cu(2/0)    & 3  & .6870  & 0.3209 \\
Cu(2/2)    & 2  & .6766  & $-0.3636$ \\
O(1/0)     & 2  & .1427  & 1.2550 \\
O(2/1)     & 2  & .1390  & 1.2214  \\
O(3/0)     & 1  & .0896  & 0.6501 \\
O(3/2)     & 1  & .0815  & 0.6384  \\
\end{tabular}
\label{asd}
\end{table}

\begin{table}
\caption{Atomic spin densities and D-values at different planar sites in the
cluster Cu$_{5}$O$_{26}$/Cu$_{8}$La$_{34}$ as obtained for spin 
multiplicities $m = 6, 4,$ and 2 with the DF method.
}
\begin{tabular}{crrrrrr}
           & \multicolumn{2}{c}{m = 6} & \multicolumn{2}{c}{m = 4} & 
                                \multicolumn{2}{c}{m = 2} \\
position   & $\rho$ &      D & $\rho$ &  D  & $\rho$ &  D \qquad \\ \hline
Cu(0/0)    & .6745  & 1.1963 & $-.2636 $& 3.2337 & $-.3036$ &   1.5959 \\
Cu(2/0)    & .6743  &$-0.9500$& .5738 & $-1.7329$ &  .5412  & $-1.6747$ \\
O(1/0)     & .1356  & 1.2034 & .0260 &    0.3062  &  .0199  &   0.2086 \\
O(3/0)     & .0818  & 0.6258 & .0726 &    0.5500  &  .0677  &   0.5228 \\
O(2/1)     & .0872  & 0.6417 & .0705 &    0.5603  &  .0660  &   0.4520 \\
Cu(0/2)    &        &        &       &            & $-.6364$  & 1.3237 \\
Cu(0/$-2$) &        &        &       &            & $.5310$ & $-1.6093$ \\
O(0/1)     &        &        &       &            & $-.0893$ & $-0.8956$ \\
O(0/3)     &        &        &       &          & $-.0778$ & $-0.6023$ \\
O(0/$-1$)  &        &        &       &          & $ .0143$ & $ 0.3083$ \\
O(0/$-3$)  &        &        &       &          & $ .0697$ & $ 0.5043$ \\
O(1/2)     &        &        &       &          & $-.0773$ & $-0.5994$ \\
O(2/$-1$)  &        &        &       &          & $-.0651$ & $0.5345$ \\
O(1/$-2$)  &        &        &       &          &  $.0657$ & $0.5265$ \\
\end{tabular}
\label{asdm}
\end{table}

\begin{table}[h,b] \centering
\caption {Values for the dipolar hyperfine couplings for Cu and O(p) for different clusters.
}
\begin{tabular}{lcccc}
cluster          & \multicolumn{2}{c}{${\rm a_{dip}^{\parallel}}$(Cu)}  
			&\multicolumn{2}{c}{${\rm c_{dip}^{\parallel}}$(O)} \\
\hline
                                    &   HF      &  DF        &   HF   &   DF \\
CuO$_{6}$/Cu$_{4}$La$_{10}$         & $-4.421$  & $-3.526$   &   .181 & .388 \\
Cu$_{2}$O$_{11}$/Cu$_{6}$La$_{12}$  & $-4.430$  & $-3.493$   &   .322 & .648 \\
Cu$_{4}$O$_{20}$/Cu$_{8}$La$_{26}$  & $-4.452$  & $-3.479$   &   .322 & .656 \\
Cu$_{5}$O$_{26}$/Cu$_{8}$La$_{34}$  & $-4.486$  & $-3.249$   &   .324 & .668 \\
Cu$_{9}$O$_{42}$/Cu$_{12}$La$_{40}$ & $-4.491$  & $-3.370$   &   .324 & .684  \\
\end{tabular}
\label{tab:adip}
\end{table}

\begin{table}
\caption{
Compilation of results for the hyperfine interactions at O(p) and 
comparison with values extracted from experiments.}
\begin{tabular}{lccccccc}
                   &  c    & ${\rm c_{dip}^{xx}}$ & ${\rm c_{dip}^{yy}}$ & ${\rm c_{dip}^{zz}}$ &
 ${\rm c_{tot}^{xx}}$ & ${\rm c_{tot}^{yy}}$ & ${\rm c_{tot}^{zz}}$ \\  \hline 
present            & 0.648 &  0.369 & -0.177 & -0.192 & 1.017 & 0.471 & 0.456  \\
La$_2$CuO$_4$ $^a$ &       &        &        &        & 0.748 & 0.480 & 0.527  \\
La$_2$CuO$_4$ $^b$ &       &        &        &        & 0.863 & 0.511 & 0.615  \\
\end{tabular}
\label{tab:compilo}
$^a$ Ref. \onlinecite{zha}\\
$^b$ Ref. \onlinecite{russ1}\\
\end{table}

\begin{table}[h,b] \centering
\caption {Contributions to the core-polarization at an oxygen, ${\rm D(O)}$, with 
one (N=1) adjacent Cu ion, 
between two (N=2) adjacent Cu ions, and difference. The numbers in parentheses are 
contributions from AOs centered at neighboring nuclei.
}
\begin{tabular}{lrrrrrr}
atomic orbital  &  \multicolumn{2}{c}{N = 1}  &\multicolumn{2}{c}{N = 2} 
& \multicolumn{2}{c}{Difference}  \\\hline
1s              &  $-0.141$& (.000)  & $-0.229$& (.001)   & $-0.088$ &(.001) \\
2s              &  $ 0.772$& (.012)  & $ 1.503$& (.052)   & $ 0.731$ &(.041) \\ \hline
Total           &  $ 0.632$& (.012)  & $ 1.274$& (.053)   &   0.643  &(.042)     \\
\end{tabular}
\label{tab:fermio}
\end{table}

\begin{table}[h,b] \centering
\caption {Contributions to the core-polarization at a copper, D(Cu), with no adjacent Cu ions (N=0),
with one NN Cu ion, and their difference. The numbers in parentheses are 
contributions from AOs centered at neighboring nuclei.
}
\begin{tabular}{lrrrrrr}
atomic orbital  &  \multicolumn{2}{c}{N = 0}  &\multicolumn{2}{c}{N = 1} 
& \multicolumn{2}{c}{Difference}  \\\hline
1s              &  $-0.060$& (.000)  & $-0.059$& (.000)   & $0.001$ &(.000) \\
2s              &  $-3.733$& (.001)  & $-3.687$& (.002)   & $0.046$ &(.001) \\
3s              &  $ 2.659$& (.009)  & $ 2.618$& (.009)   & $-0.041$&(.000) \\
4s              &  $-0.650$& (.001)  & $-0.029$& ($-.003$)&  0.621  &$(-.004)$\\ \hline
Total           &  $-1.784$& (.011)  & $-1.157$&  (.008)  &  0.627  &($-.003)$     \\
\end{tabular}
\label{tab:fermicu}
\end{table}

\begin{table}
\caption{
Compilation of results for the hyperfine interactions at Cu and comparison with values
extracted from experiments.}
\begin{tabular}{lcccccccr}
 & a     &  b  & ${\rm a_{dip}^{\parallel}}$ & ${\rm b_{dip}^{\parallel}}$
& ${\rm a_{so}^{\parallel}}$ & ${\rm a_{so}^{\perp}}$ & ${\rm
 a_{tot}^{\parallel}}$ & ${\rm a_{tot}^{\perp} \;}$ \\ \hline
present   & $-1.778$ & 0.714  & $-3.644$ & 0.071 & (2.405) & (0.427) & $-3.017$   & 0.471 \\
La$_2$CuO$_4$ $^a$ &   &  0.577   &       &         &         &        & $-2.955$ & 0.288 \\
La$_2$CuO$_4$ $^b$ &     & 0.607 &           & &       &       &$-1.550$ & 0.447 \\
YBa$_2$Cu$_3$O$_7$ $^c$ & $-2.050$& 0.744 & $-3.672$  & & 2.086 & 0.289 &$-3.636$ &$-0.075$\\
YBa$_2$Cu$_3$O$_7$ $^d$ &         & 0.585 &           & &       &       &$-3.115$ & 0.400 \\
YBa$_2$Cu$_3$O$_7$ $^e$ &         & 0.712 &           & &       &       &$-3.010$ &$-0.179$ \\
YBa$_2$Cu$_3$O$_7$ $^a$ &         & 0.687 &           & &       &       &$-2.748$ & 0.495 \\
YBa$_2$Cu$_3$O$_7$ $^b$ &         & 0.623 &           & &       &       &$-2.524$ & 0.591 \\
\end{tabular}
\label{tab:compilcu}
$^a$ Ref. \onlinecite{zha}\\
$^b$ Ref. \onlinecite{imai}\\
$^c$ Ref. \onlinecite{milarice}\\
$^d$ Ref. \onlinecite{mps}\\
$^e$ Ref. \onlinecite{ww}\\
\end{table}


\begin{thebibliography}{99}

\bibitem{brinkmann}
        For a review, see D. Brinkmann and M. Mali, {\em NMR Basic Principles
        and Progress\/}, (Springer, Heidelberg, 1994), Vol 31, p 171.

\bibitem{xyz}	see, e.g., A. Rigamonti, F. Borsa, and P. Carretta, 
	Rep. Prog. Phys. {\bf 61}, 1367 (1998).

\bibitem{milarice} F. Mila and T. M. Rice,
        Physica C {\bf 157}, 561 (1989).

\bibitem{mps} H. Monien, D. Pines, and C. P. Slichter, 
        Phys. Rev. B {\bf 41}, 11120 (1990).

\bibitem{shastry} B. S. Shastry,
        Phys. Rev. Lett. {\bf 63}, 1288 (1989).

\bibitem{ww} R. E. Walstedt and W. W. Warren,
		Science {\bf 248}, 1082 (1990).

\bibitem{mmp} A. Millis, H. Monien, and D. Pines, 
        Phys. Rev. B {\bf 42}, 167 (1990).

\bibitem{moni} H. Monien, P. Monthoux, and D. Pines, 
        Phys. Rev. B {\bf 43}, 275 (1991).

\bibitem{bp} V. Barzykin and D. Pines,
        Phys. Rev. B {\bf 52}, 13585 (1995).

\bibitem{aeppli} T. E. Mason, A. Schr\"oder, G. Aeppli, H. A. Mook, S. M. Hayden,
        Phys. Rev. Lett. {\bf 77}, 1604 (1996).

\bibitem{zha} Y. Zha, V. Barzykin, and D. Pines, 
        Phys. Rev. B {\bf 54}, 7561 (1996).

\bibitem{das} T. P. Das in ``Electronic Properties of Solids Using Cluster
  Methods'', edited by T.A. Kaplan and S.D. Mahanti, Plenum Publishing Corporation.

\bibitem{das1}
        N. Sahoo, S. Markert, T. P. Das, and K. Nagamine, 
        Phys. Rev. B {\bf 41}, 220 (1990).

\bibitem{das2}
        S. B. Sulaiman, N. Sahoo, T. P. Das, and O. Donzelli,
        Phys. Rev. B {\bf 45}, 7383 (1992).

\bibitem{schwarz1}
        K. Schwarz, C. Ambrosch-Draxl, and P. Blaha, Phys. Rev. B {\bf 42},
         2051 (1990).

\bibitem{schwarz2}
        C. Ambrosch-Draxl, P. Blaha, and K. Schwarz, Phys. Rev. B {\bf 44},
        5141 (1991).

\bibitem{winter}
        N. W. Winter, C. I. Merzbacher, and C. E. Violet, Appl. Spec. Rev.
        {\bf 28}, 123 (1993).

\bibitem{huesser} P. H\"usser, E. P. Stoll, H. U. Suter, and P. F. Meier,
  Physica C {\bf 294}, 217 (1998).

\bibitem{suter1} H. U. Suter, E. P. Stoll, P. H\"usser, S. Schafroth, and P. F. Meier,
	  Physica C {\bf 282-287}, 1639 (1997).

\bibitem{sulaiman}
        S. B. Sulaiman, N. Sahoo, T. P. Das, and O. Donzelli,
        Phys. Rev. B {\bf 42}, 7082 (1991).

\bibitem{martinhay} R. L. Martin and P. J. Hay,
        J. Chem. Phys. {\bf 98}, 8680 (1993).

\bibitem{martin} R. L. Martin, Phys. Rev. Lett. {\bf 75}, 744 (1995).

\bibitem{suter2} H. U. Suter, P. H\"usser, E. P. Stoll,  S. Schafroth, and P. F. Meier,
	  Hyperfine Interactions, in press.

\bibitem{lattice}   {\em Copper Oxide Superconductors},
        Ch. P. Poole, T. Datta and H.A. Farach, Wiley-Intescience Publication.

\bibitem{Wachters}  A. J. H. Wachters,
  	J. Chem. Phys. {\bf 52}, 1033 (1970).

\bibitem{pople}  R. Krishnan, J. S. Binkley, R. Seeger, and J. A. Pople,
  	J. Chem. Phys. {\bf 72}, 650 (1980).


\bibitem{g98}
 M. J. Frisch et al., Gaussian 98, Revision A.5, (Gaussian, Inc., Pittsburgh PA, 1998).

\bibitem{vwn1980}
	 S. H. Vosko, L. Wilk and M. Nussair,
	 Can. J. Phys. {\bf{58}}, 1200 (1980).

\bibitem{becke1}
	 A. D. Becke,  Phys. Rev. A {\bf{38}}, 3098 (1988).

\bibitem{becke2}  A. Becke, J. Chem. Phys. {\bf 88}, 2547 (1988).

\bibitem{LYP}  C. Lee, W. Yang, and R. G. Parr,
  Phys. Rev. B {\bf 37}, 785 (1988).

\bibitem{PW91}  J. P. Perdew and W. Yang,
  Phys. Rev. B {\bf 45}, 13244 (1992).

\bibitem{gamessus}
	 M.W. Schmidt, K.K. Baldridge, J.A. Boat, S.T. Elbert, M.S. Gordon, 
	J.H. Jensen, S. Koseki, M. Matsunaga, K.A. Nguyen, S.J. Su, 
	T.L. Windus, M. Dupuis, and J.A. Montgomery, 
	J. Comp. Chem. {\bf{14}}, 1347 (1993).

\bibitem{stoll97}
	 E. P. Stoll, unpublished.

\bibitem{imaisl}  T. Imai, C.P. Slichter, K. Yoshimura, and K. Kosage,
  Phys. Rev. Lett. {\bf 70}, 1002 (1993).

\bibitem{Bleaney}  B. Bleaney, K. D. Bowers and M. H. L. Pryce,
  Proc. R. Soc. London, Ser. A {\bf 228}, 166 (1955).

\bibitem{pickett} W. E. Pickett,
	Rev. Mod. Phys. {\bf 61}, 433 (1989).

\bibitem{haas} K. C. Haas,
	Solid State Phys. {\bf 42}, 213 (1989).

\bibitem{sternheimer} R. M. Sternheimer, Z. Naturforsch. {\bf 41a}, 35 (1985).

\bibitem{stein} J. Stein, S. B. Sulaiman, N. Sahoo, T. P. Das, Hyperfine
  Interactions {\bf 60}, 849 (1990).

\bibitem{ishida} K. Ishida, Y. Kitaoka, G.-Q. Zheng, and K. Asayama,
        J. Phys. Soc. Jpn. {\bf 60}, 3516 (1991).

\bibitem{russ1}  R. E. Walstedt, B. S. Shastry, and S-W. Cheong,
  Phys. Rev. Lett. {\bf 72}, 3610 (1994).

\bibitem{kohn}  P. Hohenberg and W. Kohn,
  Phys. Rev. {\bf 136}, B864 (1964).

\bibitem{kohnsham}  W. Kohn and L. J. Sham,
  Phys. Rev. {\bf 140}, A1133 (1965).

\bibitem{engels88} These effects have been studied in great detail for atoms and small 
	molecules of light elements by quantum chemistry calculations (see, e.g., 
	B. Engels and S. D. Peyerimhoff, J. Phys. B {\bf 21}, 3459 (1988)).

\bibitem{imai}  T. Imai, J. Phys. Soc. Japan {\bf 59}, 2508 (1990).

\end{thebibliography}
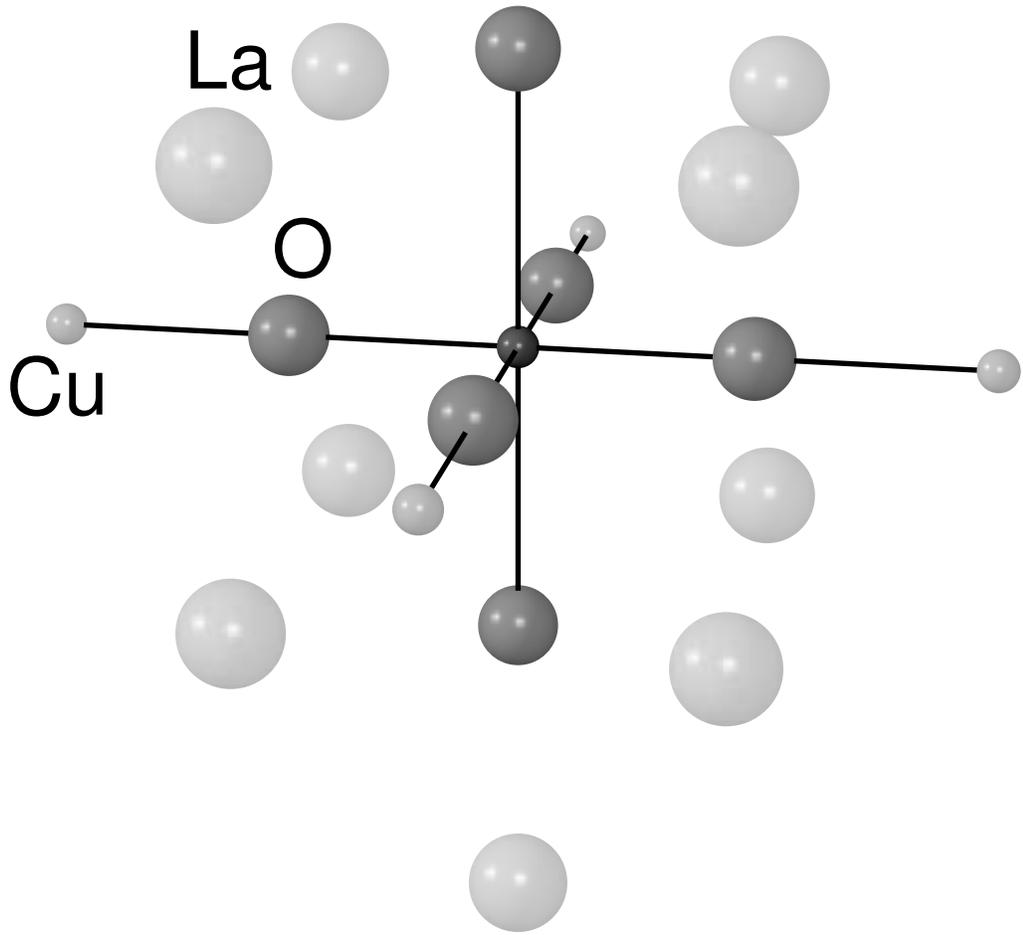
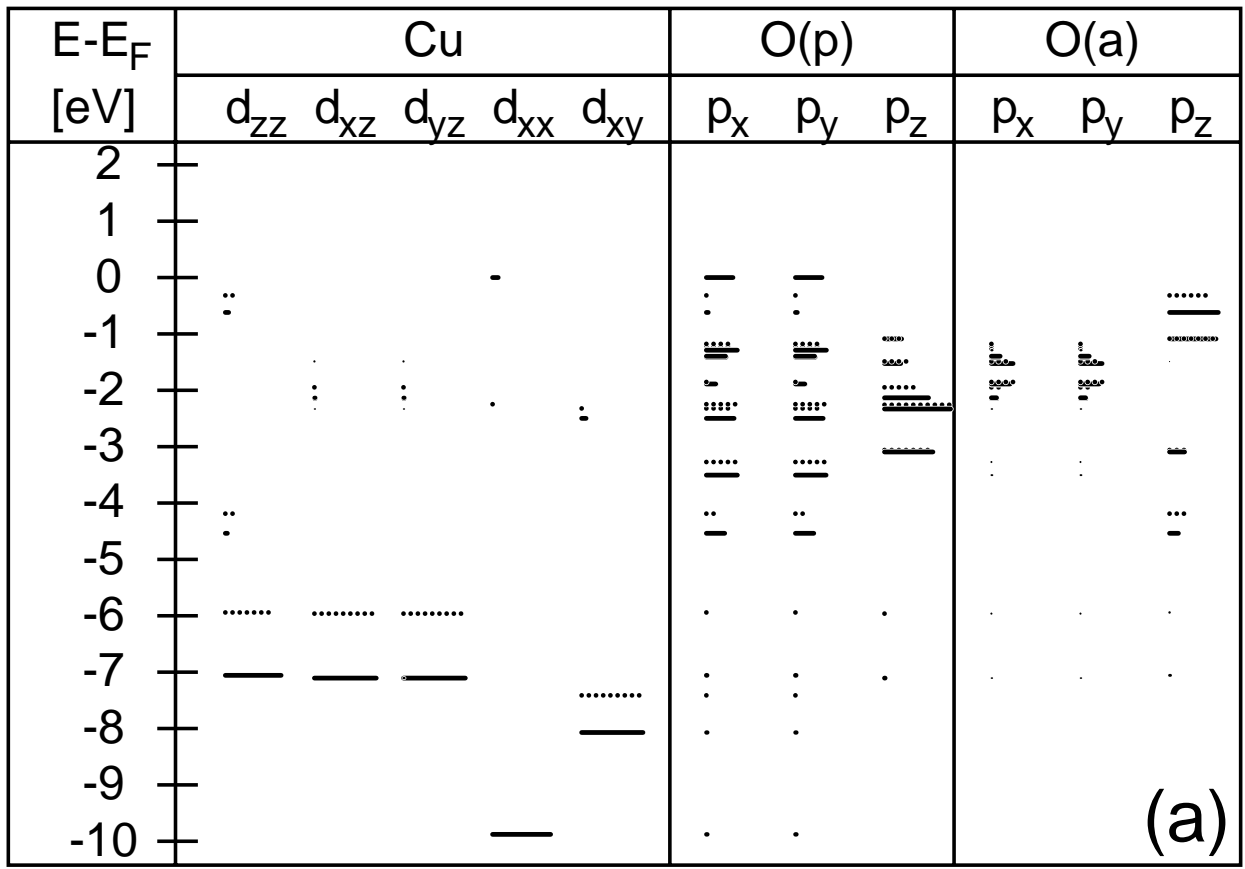
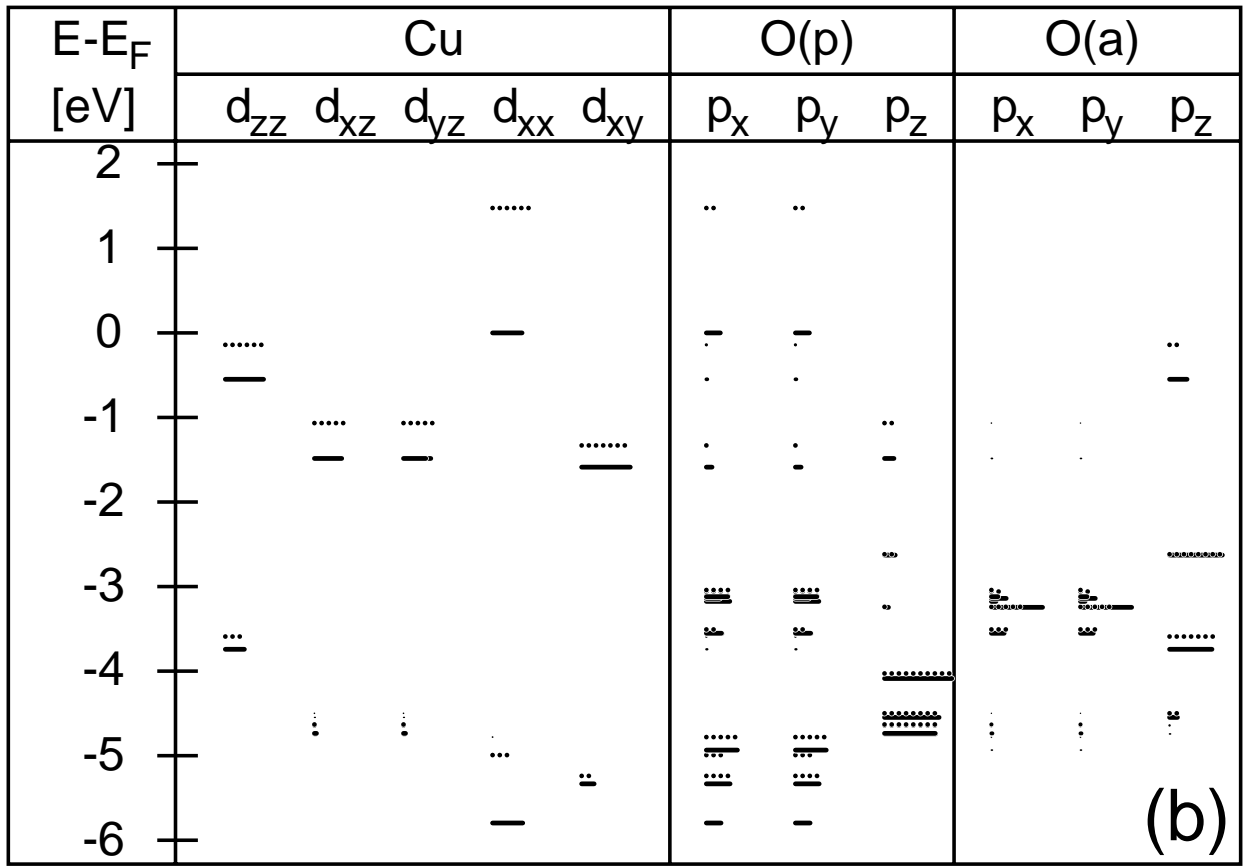
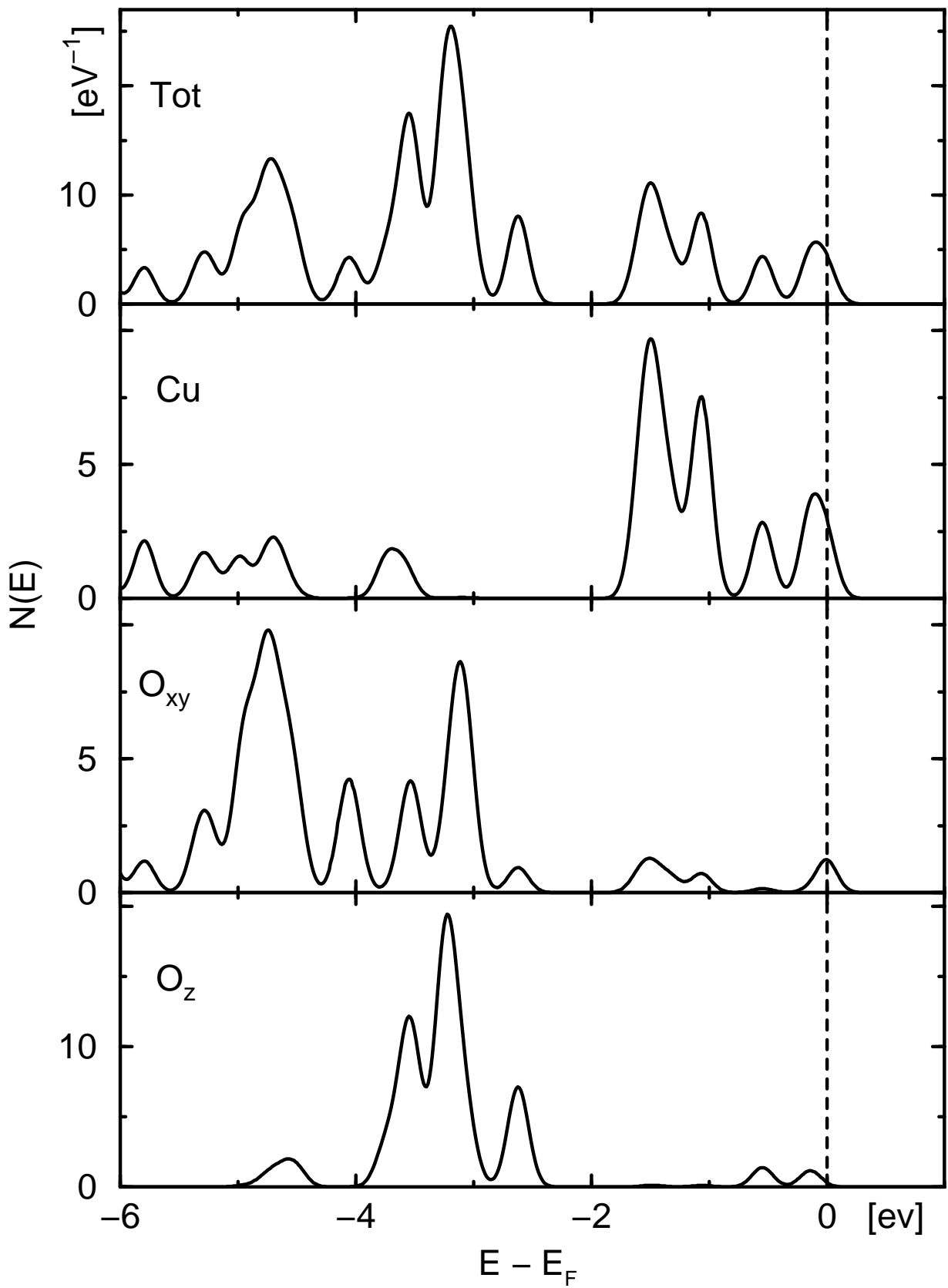
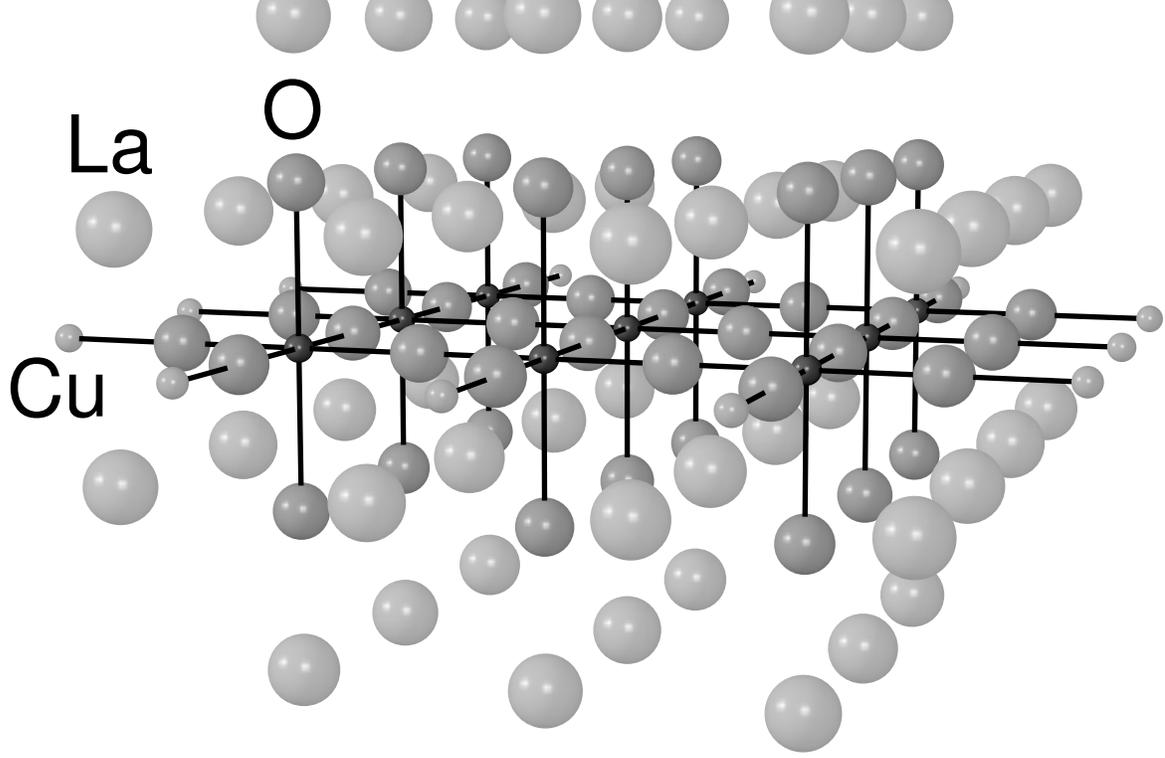
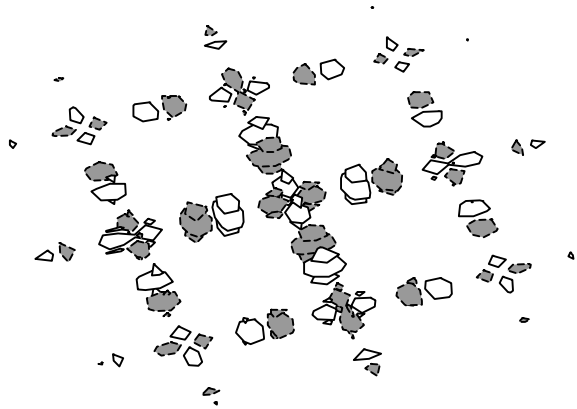
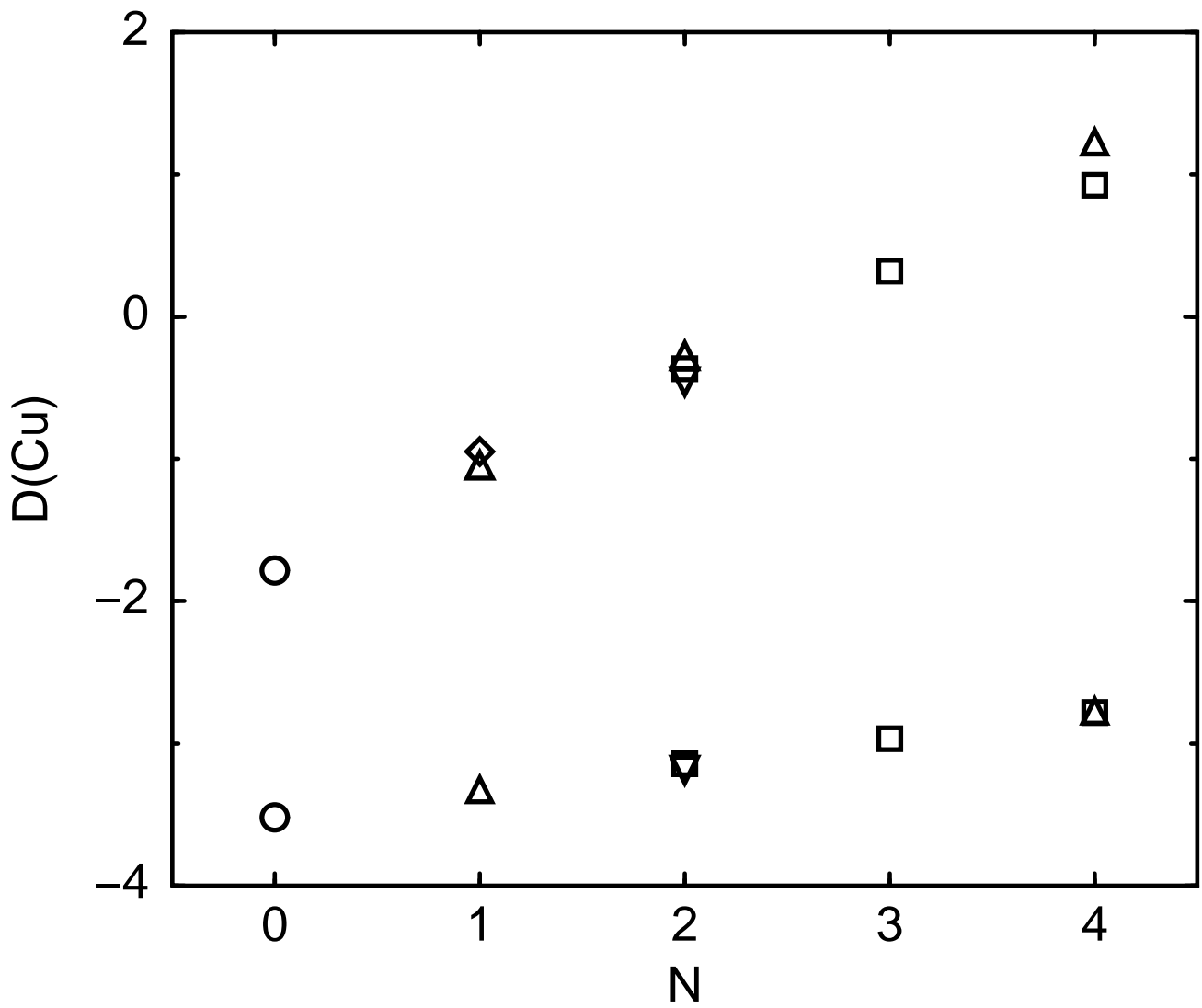
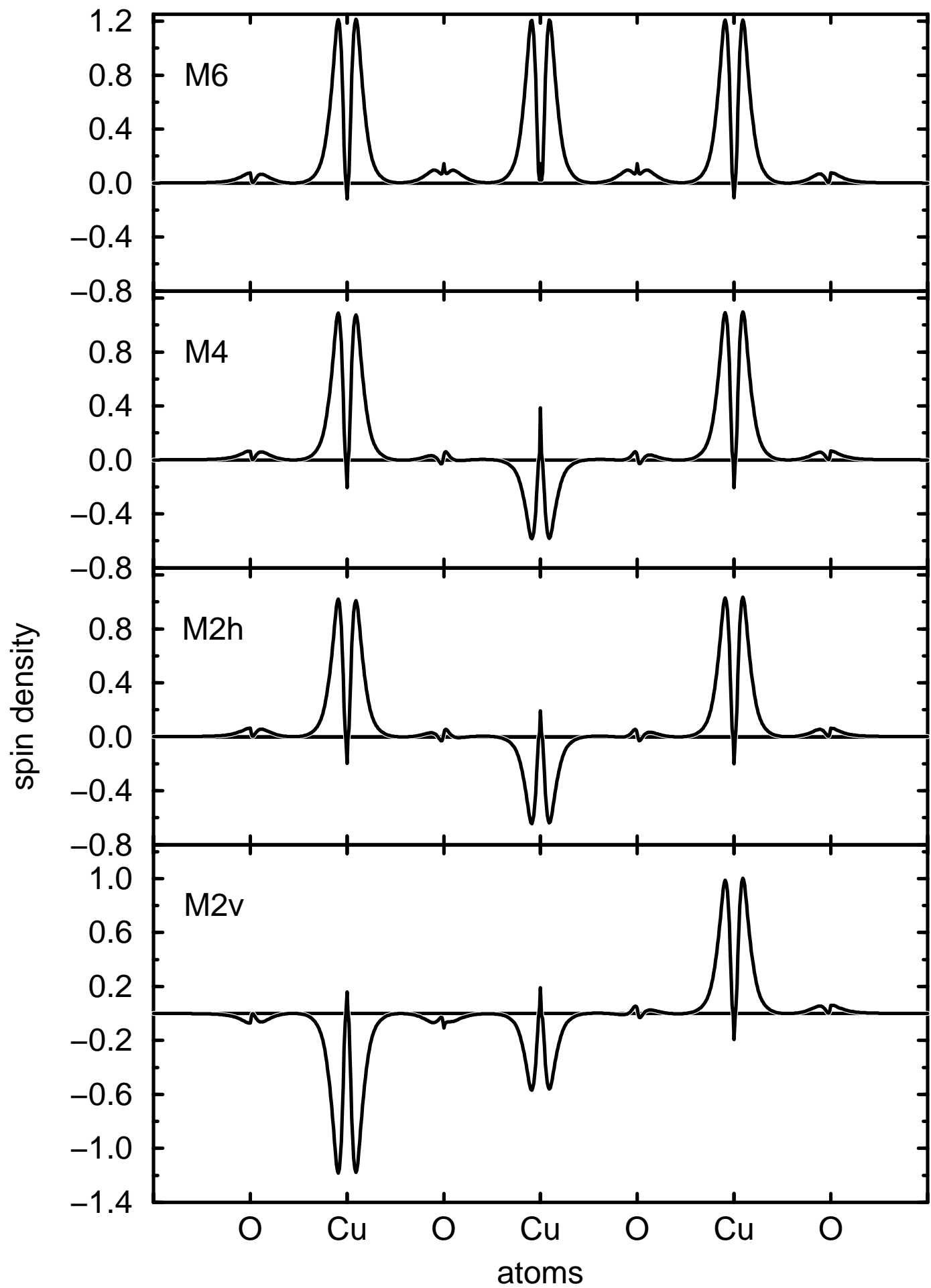
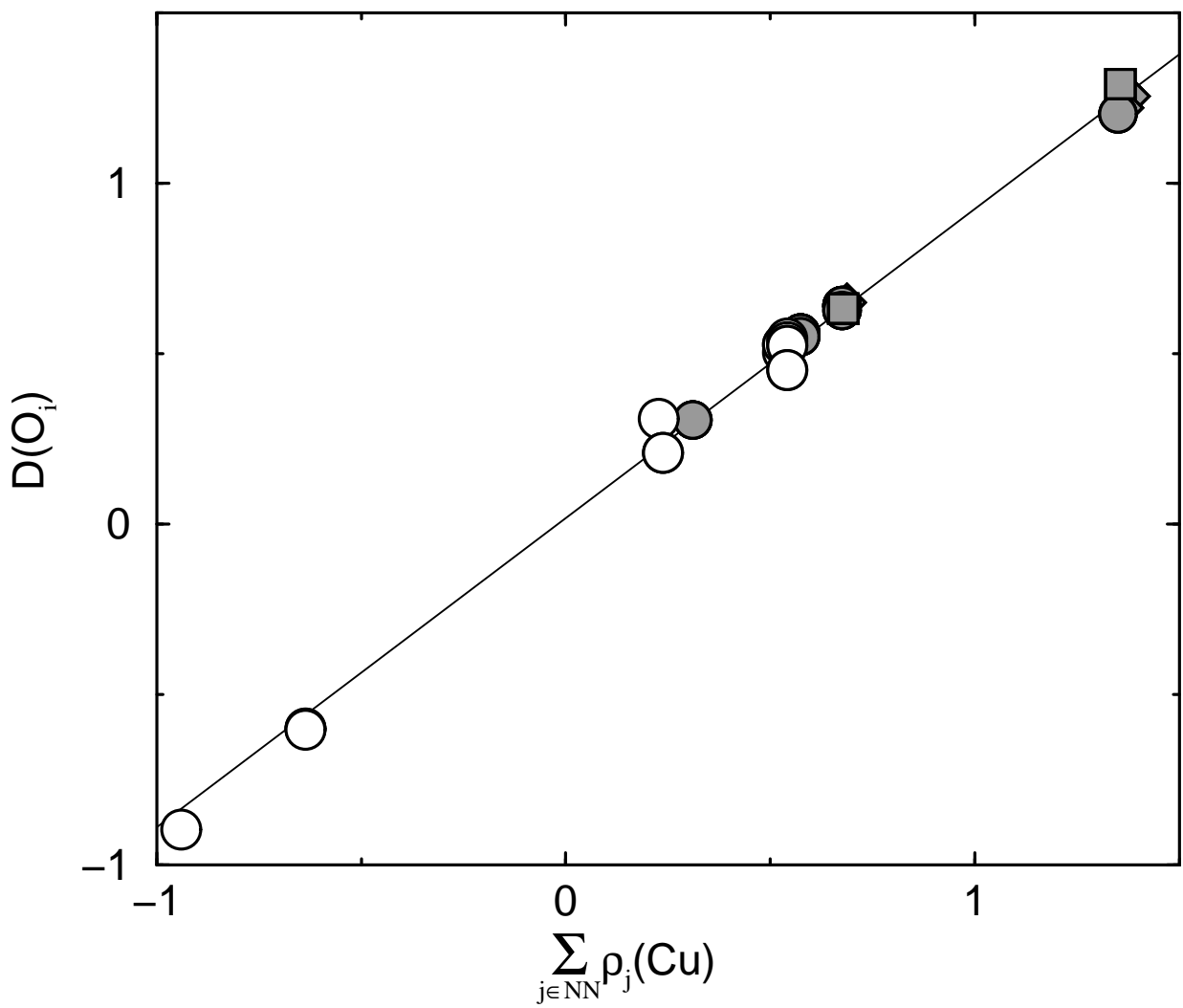
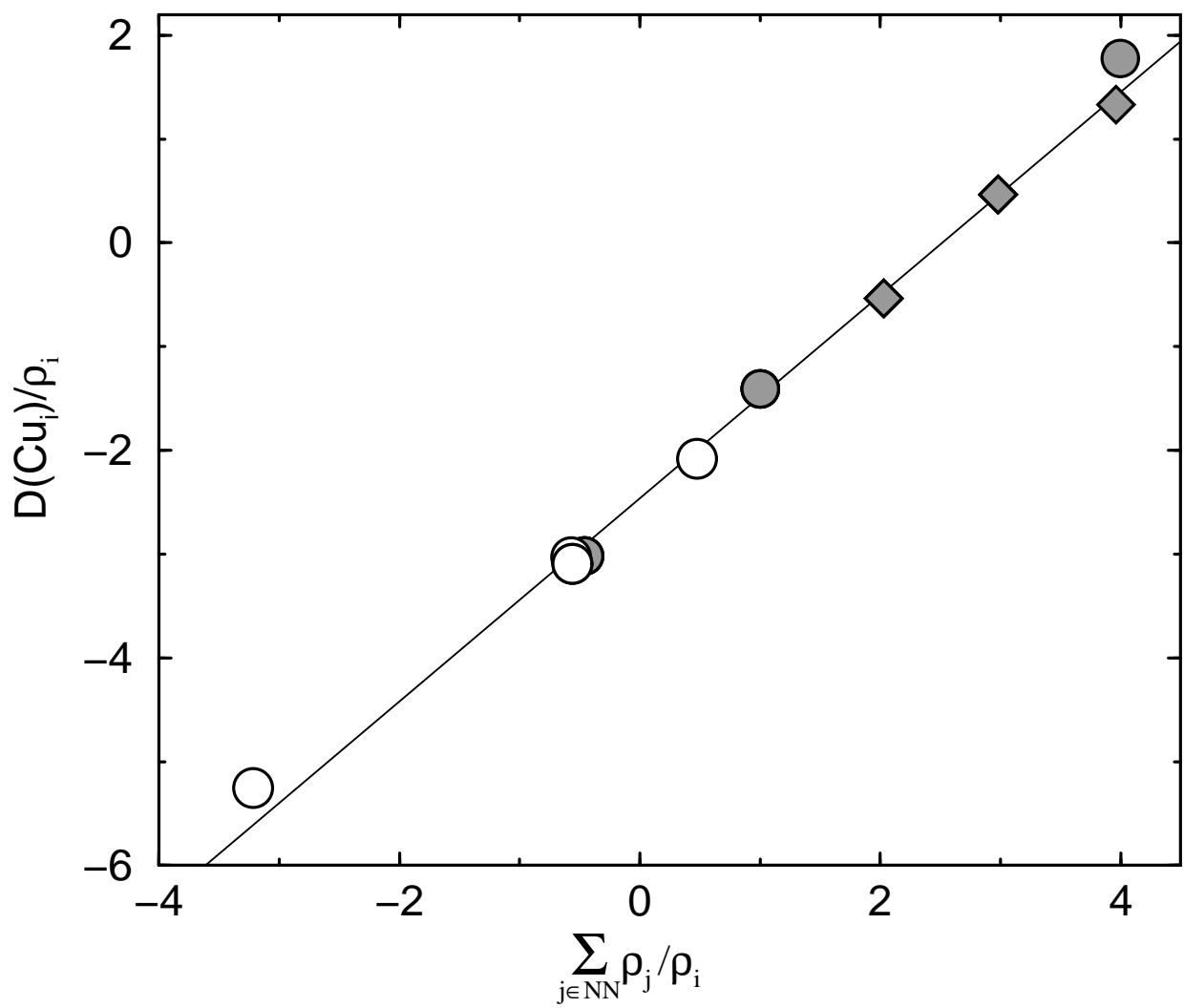
\end{document}